\documentclass[a4paper,fleqn,usenatbib]{mnras}

\usepackage{newtxtext,newtxmath}
\usepackage[T1]{fontenc}
\usepackage{ae,aecompl}

\usepackage{graphicx}	% Including figure files
\usepackage{amsmath}	% Advanced maths commands
\usepackage{amssymb}	% Extra maths symbols

\usepackage{bm}         % bold maths
\usepackage{comment}    % comment sections
\usepackage{wasysym}     % for solar system symbols
\newcommand{\AU}[0]{\mathrm{AU}}
\newcommand{\dd}[0]{\mathrm{d}}
\newcommand{\pfrac}[2]{\left( \frac{#1}{#2} \right)}
\newcommand{\brac}[1]{\left\langle #1 \right\rangle}
\newcommand{\rH}[0]{r_{\mathrm{H}}}
\newcommand{\rB}[0]{r_{\mathrm{B}}}
\newcommand{\lJ}[0]{\ell_{\mathrm{J}}}
\newcommand{\vK}[0]{v_{\mathrm{K}}}
\def\ba{\begin{eqnarray}}
\def\ea{\end{eqnarray}}

\graphicspath{ {./} }

\title[Envelopes of embedded super-Earths I]{Envelopes of embedded super-Earths\\I. Two-dimensional simulations}

\author[W. B\'ethune \& R. R. Rafikov]{
William B\'ethune$^{1}$\thanks{E-mail: wb288@damtp.cam.ac.uk},
Roman R. Rafikov$^{1,2}$
\\
$~^{1}$ Department of Applied Mathematics and Theoretical Physics, University of Cambridge, \\
Centre for Mathematical Sciences, Wilberforce Road, Cambridge CB3 0WA, UK.\\
$~^{2}$ Institute for Advanced Study, Einstein Drive, Princeton, NJ 08540
}

\date{Accepted XXX. Received YYY; in original form ZZZ}

\pubyear{2019}

\begin{document}
\label{firstpage}
\pagerange{\pageref{firstpage}--\pageref{lastpage}}
\maketitle

\begin{abstract}
Measurements of exoplanetary masses and radii have revealed a population of massive super-Earths --- planets sufficiently large that, according to one dimensional models, they should have turned into gas giants. To better understand the origin of these objects, we carry out hydrodynamical simulations of planetary cores embedded in a nascent protoplanetary disk. In this first paper of a series, to gain intuition as well as to develop useful diagnostics, we focus on two-dimensional simulations of the flow around protoplanetary cores. We use the \textsc{pluto} code to study isothermal and adiabatic envelopes around cores of sub- to super-thermal masses, fully resolving the envelope properties down to the core surface. Owing to the conservation of vortensity, envelopes acquire a substantial degree of rotational support when the core mass increases beyond the thermal mass, suggesting a limited applicability of one-dimensional models for describing the envelope structure. The finite size of the core (relatively large for super-Earths) also controls the amount of rotational support in the entire envelope. Steady non-axisymmetric shocks develop in the supersonic envelopes of high-mass cores, triggering mass accretion and turbulent mixing in their interiors. We also examine the influence of the gas self-gravity on the envelope structure. Although it only weakly alters the properties of the envelopes, the gas gravity has significant effect on the properties of the density waves triggered by the core in the protoplanetary disk. 
\end{abstract}

\begin{keywords}
  planets and satellites: gaseous planets, formation -- hydrodynamics -- methods: numerical
\end{keywords}

%%%%%%%%%%%%%%%%%%%%%%%%%%%%%%%%%%%%%%%%%%%%%%%%%%

%%%%%%%%%%%%%%%%% BODY OF PAPER %%%%%%%%%%%%%%%%%%

%%%%%%%%%%%%%%%%%%%%%%%%%%%%%%
%%%%%%%%%%%%%%%%%%%%%%%%%%%%%%

\section{Introduction}
\label{sect:intro}

%%%%%%%%%%%%%%%%%%%%%%%%%%%%%%

The Kepler survey has revealed that about $50\%$ of Sun-like stars host at least one planet with radius between $1.25 - 4 \mathrm{R}_{\oplus}$ and orbital period less than $145$ days --- the so-called super-Earths \citep{batalha13,fressin13}. Measurements of both the mass and radius of these exoplanets show that their density decreases as their size increases beyond $1.5 \mathrm{R}_{\oplus}$ \citep{weissmarcy14}. This density decrease is associated with the presence of gaseous atmospheres \citep{rogers15}, with volatile masses reaching several percent of the mass of the solid core \citep{lopezfortney14,wolfganglopez15}. 

In the standard paradigm of planet formation \citep{safronov69,kusaka70}, solid cores of protoplanets grow by accreting planetesimals \citep{nakagawa83,wetherill89} and, possibly, pebble-sized grains \citep{ormelklahr10,lambrechts12}. This scenario is supported by the observed distributions of exoplanet masses and radii \citep{matsuo07,howard12,mordasini12}. While embedded in a circumstellar disk, the growing core is surrounded by a dense and extended envelope \citep{perricameron74,rafikov06}. The observed atmospheres are most likely remnants of such primitive envelopes, modified during later stages of photoevaporation \citep{owenjackson12,owenwu13}, impact erosion \citep{Inamdar15,Yalin18} and outgassing \citep{chachanstevenson18}.

As the core grows by accretion of solids, the mass of its hydrostatic envelope increases faster than the mass of the core. Quasi-static envelopes contract and accrete gas as fast as radiative cooling permits \citep{leechiang15}. Hydrostatic and thermal equilibrium cannot be sustained beyond a critical core mass, triggering a runaway cooling and accreting phase \citep{mizuno78,pollack96}. At orbital separations $0.1 - 1\AU$, the ``runaway gas accretion'' phase is triggered above a critical core mass $5 - 20 \mathrm{M}_{\oplus}$ \citep{rafikov06}, even for vigorously accreting cores. Higher-mass cores are expected to turn into gas giants over the lifetime of the disk \citep{rafikov11}, according to one-dimensional models. Owing to their masses, many super-Earths should thus have turned into gas giants \citep{leechiang14}. That this did not happen is a puzzle. 

Various mechanisms have been proposed to moderate or prevent the runaway growth of gaseous envelopes. Examples include the ambient pressure drop following disk clearing \citep{owenwu13,ginzburg16}, the tidal (or internal) heating of close-in atmospheres \citep{ginzburg17,ginzburg18} or the enhanced opacity of dust-enriched disks \citep{leechiang14}. Hydrodynamic mechanisms have also been studied, primarily under simple thermodynamic assumptions. Due to conservation of the potential vorticity (alias vortensity), two-dimensional envelopes should become rotationally supported \citep{miki82}. If rotational support becomes dominant, a circumplanetary disk should form and regulate gas accretion \citep{rivier12,tanigawa12}. The formation of such disks depends on the efficiency of radiative cooling against heating processes, either viscous \citep{ayliffe09} or shock-induced \citep[][]{szulagyi16}.

\cite{ormel2} pointed out the possible importance of atmospheric \emph{recycling} --- dynamic exchange of mass between the envelope and the surrounding disk --- for preventing the transition of the super-Earths into a gas giant. A notable feature of their model is the inclusion of a spatially resolved inner boundary in the simulations domain, mimicking the presence of a solid core. Most previous studies have represented the core via a smoothed gravitational potential \citep[e.g., ][]{dangelo13} and/or a mass sink \citep[e.g., ][]{machida10}, focusing on the properties of the flow on the scale of the disk thickness. However, \cite{ormel1} showed the influence of this boundary on the global properties of two-dimensional planetary envelopes. 

This paper is the first in a series of numerical studies of super-Earth atmospheres, with increasing degree of realism and complexity. Its goal is to go beyond the work of \cite{ormel1} and to provide detailed 2D exploration of the atmospheres of massive (super-thermal, see \autoref{sec:method}) cores, which would be capable of triggering runaway gas accretion according to 1D models \citep{mizuno80}. We investigate the role of envelope recycling, with a specific care taken of the core boundary. In doing this we pay particular attention to the balance of pressure versus rotational support in the envelope, and to the mixing properties of the flow. 

The plan of the paper is the following. We present our model and its numerical implementation in section \ref{sec:method}. Two appendices complete this section with a series of tests. Section \ref{sec:results} contains the results of isothermal, adiabatic, and finally self-gravitating simulations. Section \ref{sec:discussion} provides a discussion of our results in light of previous studies, and summarize our main conclusions in section \ref{sec:summary}.

%%%%%%%%%%%%%%%%%%%%%%%%%%%%%%%%%%%%%%%%%%%%%%%%%%%
%%%%%%%%%%%%%%%%%%%%%%%%%%%%%%%%%%%%%%%%%%%%%%%%%%%

\section{Method}
\label{sec:method}
\subsection{Physical setup}

%%%%%%%%%%%%%%%%%%%%%%%%%%%%%%%%%%%%%%%%%%%%%%%%%%%

We consider a solid core orbiting a star on a circular trajectory in the midplane of a protoplanetary disk. In this paper, we focus on the two-dimensional properties of the flow around the core, with simple assumptions about the thermodynamics of the gas. The gas self-gravity is included in some cases, see \autoref{sec:selfgravity}.

Let $r_c$ be the radius of the core, $m_c$ its mass, $a$ its semi-major axis, and $\Omega=(Gm_\star/a^3)^{1/2}$ its Keplerian frequency around the central star of mass $m_\star$. The sound speed of the gas $c_s$ is linked to the hydrostatic pressure scale height of the disk $h=c_s/\Omega$. Another important lengthscale of the problem is the Bondi radius\footnote{The Bondi radius is often defined with the Keplerian \emph{escape velocity} $\rB = 2 G m_c / c_s^2$, twice larger than our definition.} $\rB = G m_c / c_s^2$. 

We define the dimensionless ratios $H\equiv h/r_c$ and $B\equiv \rB/r_c$, and evaluate their typical values for a super-Earth core orbiting a Solar-mass star:
\begin{align}
H &\approx 60 \pfrac{a}{0.1\AU} \pfrac{h/a}{5\%} \pfrac{r_c}{2r_{\oplus}}^{-1}, 
\label{eqn:magnH}\\ %1.174
B &\approx 14 \pfrac{a}{0.1\AU} \pfrac{m_c}{10m_{\oplus}} \pfrac{r_c}{2r_{\oplus}}^{-1} \pfrac{m_{\star}}{m_{\odot}}^{-1} \pfrac{h/a}{5\%}^{-2}.
\label{eqn:magnH}
\end{align}
For an Earth-like planet at $1\AU$, both $H$ and $B$ would be much larger than these numerical estimates. However, for characteristic super-Earth-like core masses and separations from the star, $B$ can be of order unity and $r_c$ can reach several percent of $h$. As a result, the physical core size might be able to influence the flow structure on scales $\sim h$. This is an important characteristic of super-Earths that distinguishes them from e.g. the cores of gas giants at 5-10 AU.

From \eqref{eqn:magnH} and \eqref{eqn:magnH} one can form the dimensionless ratio 
\begin{align}
\frac{B}{H} =\frac{m_c}{m_{\rm th}}&\approx 0.24 \pfrac{m_c}{10m_{\oplus}} \pfrac{m_{\star}}{m_{\odot}}^{-1} \pfrac{h/a}{5\%}^{-3},
\label{eqn:magnBoH} %2.403 
\end{align}
which is independent of the core radius. This ratio is equivalent to the ratio of the core mass $m_c$ to its `thermal mass'  \citep{rafikov06}: 
\ba  
m_{\rm th}=\frac{c_s^3}{\Omega G}=m_\star\left(\frac{h}{a}\right)^3\approx 40m_\oplus\pfrac{m_{\star}}{m_{\odot}}\pfrac{h/a}{5\%}^{3}
\ea  
At this important mass scale the perturbations induced by the core in the surrounding disc (e.g., density waves) become non-linear. Also, $m_c=m_{\rm th}$ corresponds to $\rB=h$, and both scales are approximately equal to the core's Hill radius defined as $\rH=a\left(m_c/3m_\star\right)^{1/3}$ (up to a factor $3^{1/3}$).

Note that $h$ and $\rB$ are defined with respect to the temperature of the disk. If the disc is non-isothermal, then any heating process occuring near the core will increase the effective pressure scale and decrease the effective Bondi radius. In this work we use either isothermal ($P = \rho c_s^2$) or adiabatic ($P \propto \rho^{\gamma}$) equation of state for the gas. The isothermal limit would correspond to instantaneous radiative heating/cooling, while the adiabatic limit assumes that no radiative losses occur. In the adiabatic case, we prescribe the exponent $\gamma=7/5$, as would be appropriate for a diatomic gas in three dimensions (even though we use a 2D setup in this work). For simplicity, chemical effects such as thermal dissociation and reactions between different species are not taken into account. Also, unlike \cite{kley99}, we consider only inviscid flows in which momentum dissipation should be limited to shocks \citep{lubow99}.

%%%%%%%%%%%%%%%%%%%%%%%%%%%%%%%%%%%%%%%%%%%%%%%%%%%
%%%%%%%%%%%%%%%%%%%%%%%%%%%%%%%%%%%%%%%%%%%%%%%%%%%

\subsection{Numerical setup}

%%%%%%%%%%%%%%%%%%%%%%%%%%%%%%%%%%%%%%%%%%%%%%%%%%%

We aim at resolving scales ranging from a fraction of the core radius to several pressure scales $h$. Since $H \gg 1$, the envelope should isolate the core from the global dynamics of the disk. We follow the core along its orbit around the star in `the local approximation' \citep{hill78}. Let $(x,y)$ be Cartesian coordinates with the origin at the center of the core, with $x$ axis pointing from the star to the core and $y$ along the orbit of the core. The total gravitational potential $\Phi$ is split into three components: the core potential $\Phi_c$, the gas self-gravity $\Phi_g$, and the tidal potential of the star in the co-orbiting frame. In the local approximation, this decomposition reads
\begin{equation} 
\label{eqn:totpotgrav}
  \Phi = \Phi_c + \Phi_g + q\Omega^2 x^2,
\end{equation}
where $q \equiv \Omega^{-1}\partial_x v_{y,0}$ is the dimensionless shear rate of the background flow $v_{y,0}$ (unperturbed by the core). We set the shear rate to its Keplerian value $q=-3/2$ throughout this paper. 

For simplicity, we neglect the head-wind experienced by the core in sub-Keplerian disk (its effect was previously examined by \cite{ormel1}). If the midplane pressure $P\propto r^{-n}$, then the core faces a head-wind with Mach number $v/c_s \simeq (n/2)(h/a)$, i.e. a few per cent for typical values of the opening angle $h/a\approx5\%$. 

A key aspect of our model is the inclusion of the core as a physical boundary. Matter should not be able to cross the boundary, allowing the accumulation of mass and angular momentum on top of the core. However, resolving the core comes at a cost. At spatial scales smaller than the core size, hydrodynamic fluctuations evolve on time scales shorter than $r_c/c_s = (\Omega H)^{-1} \ll \Omega^{-1}$. Following the dynamics on scales $\sim r_c$ over several orbits is computationally affordable only for moderate values of $H$. Luckily, the time required to restore a quasi-static equilibrium is the sound crossing-time $\sim \Omega^{-1}$ \citep[see][and Appendix \ref{app:convstudy}]{miki82}. For this reason, we focus on time intervals of a few tens of orbits only, short relative to disk clearing or planet migration timescales \citep[][]{gorti16,funglee18}. In this sense, we look at the quasi-instantaneous state of the envelope in this work. 

We call $\rho$ the gas (surface) density, $v$ the velocity, $e$ the internal energy per unit mass, $P$ the thermal pressure, $\Phi$ the gravitational potential from \eqref{eqn:totpotgrav} and $E\equiv \rho \left(e + v^2/2 + \Phi \right)$ the total energy density. With these notations, the equations describing the dynamics of mass, momentum and energy are:
\begin{alignat}{3}
  &\partial_t \rho &&+ \nabla\cdot\left[\rho v\right] &&= 0\,,\label{eqn:consrho}\\
  &\partial_t \left[\rho v\right] &&+ \nabla\cdot\left[ \rho v\otimes v + P \,\mathbb{I} \right] &&= - \rho \nabla \Phi - 2 \rho \Omega\times v\,, \label{eqn:consrhov}\\
  &\partial_t E &&+ \nabla\cdot\left[ \left(E+P\right) v \right] &&= 0. \label{eqn:conse}
\end{alignat}
We close this system with either an isothermal ($P = \rho c_s^2$) or adiabatic ($P = \left(\gamma-1\right) \rho E$) equation of state. In the isothermal case, only \eqref{eqn:consrho} and \eqref{eqn:consrhov} are actually integrated.

%%%%%%%%%%%%%%%%%%%%%%%%%%%%%%%%%%%%%%%%%%%%%%%%%%%

\subsubsection{Integration scheme} 
\label{sec:numscheme}

%%%%%%%%%%%%%%%%%%%%%%%%%%%%%%%%%%%%%%%%%%%%%%%%%%%

We use the finite-volume code \textsc{pluto} \citep{mignone07} to integrate \eqref{eqn:consrho}$-$\eqref{eqn:conse} in conservative form. The primitive variables $\left(\rho,v,P\right)$ are estimated at cell interfaces by linear reconstruction with Van Leer's slope limiter \citep{vanleer79}. Godunov fluxes are then computed via the Roe approximate Riemann solver \citep{roe81}. A shock-flattening strategy is set to stabilize the solver in regions of strong pressure contrast. If the relative pressure variations between neighboring cells exceed a factor of $5$, we locally switch to the MINMOD slope limiter and to the HLL approximate Riemann solver \citep{van1997relation}. The time-stepping is performed via an explicit second-order Runge-Kutta scheme; a Courant number of $0.3$ is used for CFL stability. 

Specific care is taken of the Coriolis acceleration: to ensure the local conservation of angular momentum, the equations are discretized in a frame rotating with angular velocity $\Omega$ around the core axis \citep{kley98,mignone12}. When taking the gas self-gravity into account, the gas potential $\Phi_g$ is obtained by solving Poisson's equation as described in Appendix \ref{app:selfgravity}. 

%%%%%%%%%%%%%%%%%%%%%%%%%%%%%%%%%%%%%%%%%%%%%%%%%%%

\subsubsection{Computational domain}

%%%%%%%%%%%%%%%%%%%%%%%%%%%%%%%%%%%%%%%%%%%%%%%%%%%

We use polar coordinates $(r,\varphi)$ centered on the core, with $\varphi=0$ along the axis from the star to the core. The computational domain is $\left(r,\varphi\right) \in \left[r_c,128 r_c\right] \times \left[0, 2\pi \right]$. The radial interval is meshed with $512$ logarithmically spaced grid cells, and the azimuthal interval is uniformly meshed by $640$ cells. This corresponds to $73$ cells over $\left[r, 2 r \right]$ for any $r$. The convergence of most diagnostics was tested for resolutions ranging from $64$ to $512$ radial cells (see Appendix \ref{app:convstudy}). 

The logarithmic grid spacing allows a fine spatial resolution near the core at the expense of computational time. The CFL constraint is dominated by sound waves in the innermost grid cells. For moderate values of $H=16$, approximately $5\times 10^4$ time steps are required to integrate one orbit of the core; this number increases linearly with $H$ in isothermal simulations. The timestep constraint is even more demanding in adiabatic simulations because the sound speed increases near the core. We therefore focus our analysis on $H\leq 32$ simulations; according to \eqref{eqn:magnH}, this parameter range is most relevant to large cores at small orbital separations in thin disks.

%%%%%%%%%%%%%%%%%%%%%%%%%%%%%%%%%%%%%%%%%%%%%%%%%%%

\subsubsection{Initial and boundary conditions} 
\label{sec:initboundcond}

%%%%%%%%%%%%%%%%%%%%%%%%%%%%%%%%%%%%%%%%%%%%%%%%%%%

The initial conditions consist of the unperturbed shear flow of the disk $\left(\rho,v_x,v_y,P\right) = \left(\rho_0,0,-3\Omega x/2,\rho_0 c_s^2\right)$. To avoid a violent relaxation towards the core, the potential of the core $\Phi_c$ is gradually introduced, increasing linearly over the first two orbits (see Appendix \ref{app:convstudy}). 

We impose periodic boundary conditions in the azimuthal direction. In our local setup we cannot capture certain features of the global dynamics of the disk (e.g., the opening of a gap by the core) without ad-hoc prescriptions at the outer radial boundary. The shear flow is supersonic with respect to the boundary for $\left\vert x\right\vert > 2h/3$, so unless filtering out the characteristic waves that propagate inward, any density/velocity perturbation reaching the boundary should generate discontinuities inside the computational domain. For simplicity, we impose the initial conditions of an unperturbed shear flow at the outer radial boundary. Because of the sustained perturbations induced by the planet on the flow, a velocity discontinuity appears at the outermost radial grid cells in the portion of the flow leaving the computational domain. We expect no influence of this discontinuity in the inner regions close to the core. As a precaution, the outer regions $r>64 r_c$ are always be excluded from our analysis.  

At the inner radial boundary, we require that mass should not flow through it. This is achieved by the adequate symmetrization
\begin{equation}
\left[\rho,v_r,P,\Phi_c\right](r_c-x) = \left[+\rho,-v_r,+P,+\Phi_c\right](r_c+x). 
\end{equation}
The even symmetry of $\Phi_c$ across the boundary effectively cancels the gravitational acceleration at the interface. The residual mass flux through the boundary is always monitored in our simulations. 

There is not a unique acceptable boundary condition for the azimuthal velocity $v_{\varphi}$ at the core surface. This choice will likely affect the long term behavior of our simulations by injecting or extracting angular momentum from the envelope. Let $\omega\equiv \nabla\times v$ be the vorticity, and $\varpi_z \equiv \left(\omega_z + 2\Omega\right) / \rho$ be the vertical vortensity of the flow. In the inviscid case this component should remain constant in the absence of shocks \citep[see][and \autoref{sec:circarcore}]{ormel1}. We use this property to design a stress-free inner boundary. Since $v_r$ is already set by the zero-mass-flux condition, we can impose $\omega_z$ (and therefore $\varpi_z$) by setting the derivative $\partial_r v_{\varphi}$. The velocity $v_{\varphi}(r)$ in the ghost cells is estimated by first-order integration from the active domain, followed by $8$ implicit Jacobi iterations. The desired vortensity $\varpi_z$ is thus maintained to better than $1\%$ accuracy. Since vortensity is not conserved around high-mass cores (see \autoref{sec:vortensity}), we impose the azimuthally-averaged value of the vortensity on top of the core instead of the background $\varpi_0$.

We emphasize that the system \eqref{eqn:consrho}-\eqref{eqn:conse}, along with the prescribed boundary conditions, admits an infinity of steady two-dimensional solutions. \emph{In the absence of mixing}, whether viscous or turbulent, isothermal solutions can be parametrized by a radial profile of vortensity near the core, and adiabatic solutions admit the entropy profile as an additional degree of freedom. By slowly introducing the potential of the core, we are studying the class of solutions having a constant vortensity and entropy away from shocks. 

%%%%%%%%%%%%%%%%%%%%%%%%%%%%%%%%%%%%%%%%%%%%%%%%%%%

\subsubsection{Gravitational potential of the core} 
\label{sec:gravpot}

%%%%%%%%%%%%%%%%%%%%%%%%%%%%%%%%%%%%%%%%%%%%%%%%%%%

In this simplified study, we neglect the vertical dimension of the protoplanetary disk: the flow is constrained to evolve in the midplane of the disk, where the core lies. If this model is meant to represent a three-dimensional disk in some average sense, then the gravitational potential of the core $\Phi_c$ should be averaged correspondingly. 

It is customary to smooth the gravitational potential of the core in order to avoid singularities when the core size is unresolved at the grid scale \citep{muellerkleymeru12}. Let $\Phi_{\mathrm{Newton}} = -Gm_c/r$ be the Newtonian potential of a point mass $m_c$; the most common smoothing technique \citep{plummer11} includes a smoothing length $\epsilon$ via Plummer potential $\Phi_{\mathrm{Plummer}}(r) = - Gm_c/\sqrt{r^2+\epsilon^2}$. To avoid spurious mass fluxes through the core boundary of their three-dimensional simulations, \cite{ormel2} used a potential $\Phi_{\mathrm{Ormel}}$ that is force-free ($\partial_r \Phi_{\mathrm{Ormel}}=0$) near the core surface (see their equation 2). As indicated above, we prevented such mass leaks by full symmetrization of the ghost cells at the inner boundary. 

\begin{figure}%[H]
\begin{center}
\includegraphics[width=\columnwidth]{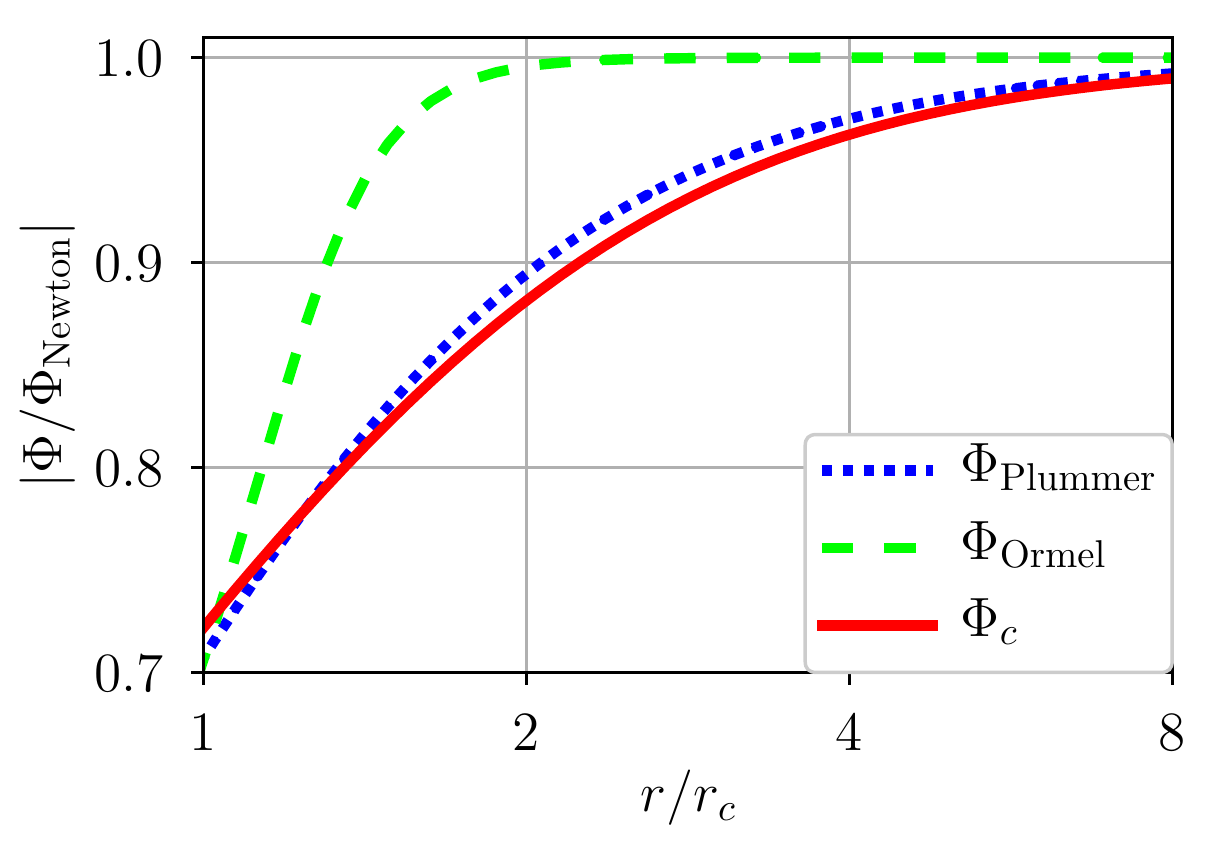}
\caption{Different softening methods for the gravitational potential of the core: Plummer potential $\Phi_{\mathrm{Plummer}}$ (dotted blue), force-free potential $\Phi_{\mathrm{Ormel}}$ (dashed green, see text), and vertically averaged potential $\Phi_c$ used in this study (solid red), normalized by the Newtonian potential $\Phi_{\mathrm{Newton}}$ of an identical mass. \label{fig:coresmooth}}
\end{center}
\end{figure}

In this work we opted to integrate the Newtonian potential $-1/\sqrt{r^2+z^2}$ over the height of the core $z\in\left[\pm r_c\right]$ to obtain the softened potential
\begin{equation} 
\label{eqn:defcorepot}
\Phi_c(r) = \frac{Gm_c}{2\epsilon} \log\left( \frac{\sqrt{r^2+\epsilon^2} - \epsilon}{\sqrt{r^2+\epsilon^2} + \epsilon}\right),
\end{equation}
with the smoothing length $\epsilon=2 r_c$. \autoref{fig:coresmooth} illustrates these different smoothing techniques. In all cases, the smoothing becomes apparent only near the core radius. Our potential $\Phi_c$ nearly coincides with $\Phi_{\mathrm{Plummer}}(\epsilon=r_c)$ for $r>r_c$ so we expect no significant difference with a Plummer type of smoothing. 

%%%%%%%%%%%%%%%%%%%%%%%%%%%%%%%%%%%%%%%%%%%%%%%%%%%

\subsubsection{Units and conventions}

%%%%%%%%%%%%%%%%%%%%%%%%%%%%%%%%%%%%%%%%%%%%%%%%%%%

From now on, we set the gravitational constant $G=1$. We take the orbital frequency $\Omega$ and the core radius $r_c$ as frequency and distance units. With this choice, the isothermal sound speed of the disk $c_s = \Omega h = H$ and the core mass $m_c = \rB c_s^2 = B\,H^2$. Without self-gravity, we take the background disk density $\rho_0$ as density unit; this value is tuned in the self-gravitating case, see \autoref{sec:selfgravity}. 

We label each simulation run by its equation of state (\verb|I| for isothermal, \verb|A| for adiabatic), its pressure scale height \verb|H|\# and its Bondi radius \verb|B|\#. Spatial averages are represented by brackets $\brac{\cdot}$, taken in the azimuthal dimension $\varphi$ by default.

%%%%%%%%%%%%%%%%%%%%%%%%%%%%%%%%%%%%%%%%%%%%%%%%%%%
%%%%%%%%%%%%%%%%%%%%%%%%%%%%%%%%%%%%%%%%%%%%%%%%%%%

\section{Results for the fiducial isothermal simulation} 
\label{sec:results}

%%%%%%%%%%%%%%%%%%%%%%%%%%%%%%%%%%%%%%%%%%%%%%%%%%%

We present our results by first examining in detail the outcomes of a single (fiducial) run. We take the run \verb|IH16B16| as such a reference isothermal simulation. With $B=H$, the core mass equals its thermal mass \citep{rafikov06}, delimiting the transition from low-mass to high-mass cores. This simulation is integrated for a total of twenty orbital periods of the core. 

\begin{figure}%[H]
\begin{center}
\includegraphics[width=\columnwidth]{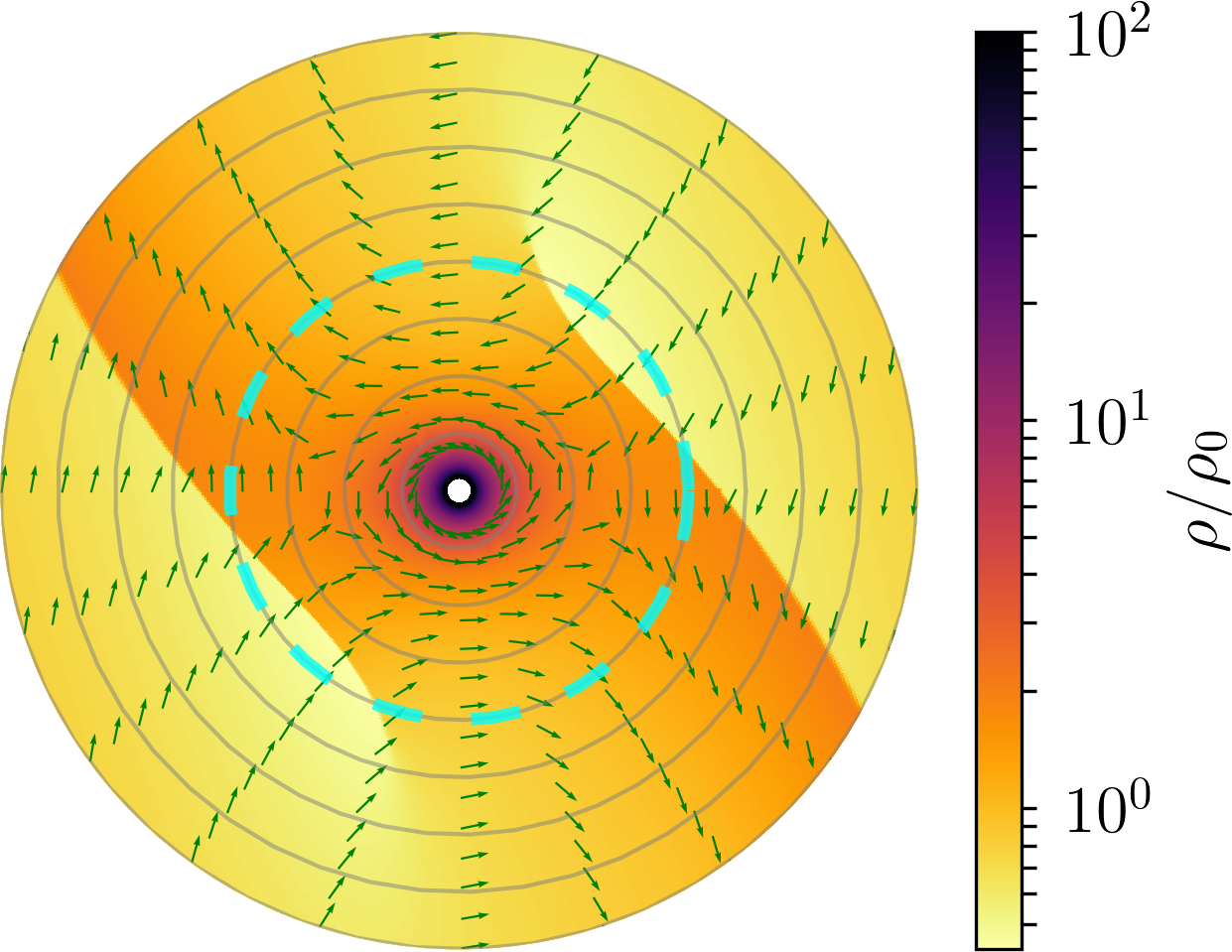}
\caption{Time-averaged flow in the reference isothermal simulation \texttt{IH16B16}: density relative to its background value $\rho_0$ (color map) and velocity field (green arrows, orientation only); the dashed cyan circle markes the pressure scale $r=h$, and the grey circles mark every $4r_c$; only the inner $r<2h$ are shown here. \label{fig:i2dp16b16_Map_rho}}
\end{center}
\end{figure}

\autoref{fig:i2dp16b16_Map_rho} shows the steady flow obtained in \verb|IH16B16| after time-averaging over $5$ orbits (ten snapshots). As described by \cite{miki82}, the flow can be partitionned into several distinct regions: the background shear flow (leftmost and rightmost parts), the co-orbiting gas on horse-shoe orbits (lower and upper parts), and closed streamlines around the core. The fluid on closed streamlines has a prograde orientation $v_{\varphi}>0$. The pressure scale height $h$ demarcates the dense envelope from the background shear flow. Spiral density waves are launched near the edge of the envelope and saturate in stationary shock waves. These shock waves extend out to the outer boundary of the computational domain. They maintain a near-parabolic shape due to the background shear; the amplitude of the density jump increases at first as a result of angular momentum injection by the planetary torque \citep{goodfikov01} before saturating at $\Delta \rho / \rho_0 \approx 2$ away from the core due to numerical dissipation.

%%%%%%%%%%%%%%%%%%%%%%%%%%%%%%%%%%%%%%%%%%%%%%%%%%%

\subsection{Radial momentum balance}

\begin{figure}%[H]
\begin{center}
\includegraphics[width=1.0\columnwidth]{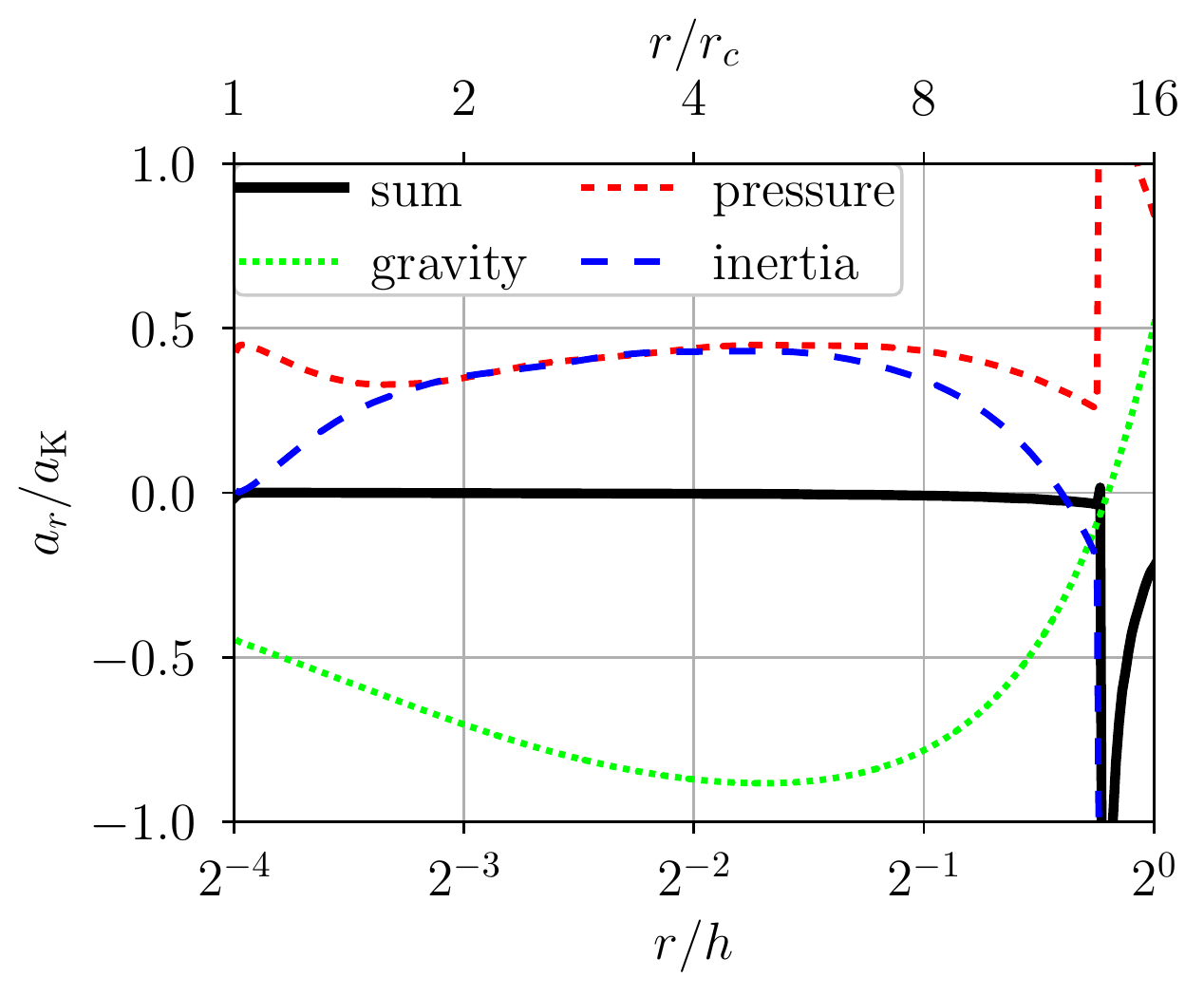}
\caption{Azimuthally-averaged radial accelerations compared to the Keplerian acceleration $a_{\mathrm{K}}=m_c/r^2$ in the fiducial run \texttt{IH16B16}; the total acceleration (solid black) is split into gravity (dotted green), pressure gradient (dashed red), and inertial terms (dashed blue). The total acceleration is close to zero in the inner part of the envelope, justifying \eqref{eqn:radmombal}. \label{fig:i2dp16b16_SH_axel}}
\end{center}
\end{figure}

In a steady and non-accreting state, the azimuthally-averaged velocity $\brac{v_r} \simeq 0$ (as supported by our numerical results), so that we can express radial momentum balance as
\begin{equation} 
\label{eqn:radmombal}
  -\underbrace{\partial_r \Phi}_\text{gravity} -\underbrace{\partial_r P / \rho}_\text{pressure} + \underbrace{\left(2\Omega + v_{\varphi}/r\right) v_{\varphi}}_\text{inertial support} = 0.
\end{equation}
These contributions are drawn on \autoref{fig:i2dp16b16_SH_axel} and compared to the Keplerian acceleration $a_{\mathrm{K}} \equiv m_c/r^2$. The gravitational acceleration is sub-Keplerian at small radii due to the smoothing of the core potential $\Phi_c$. The gravitational acceleration changes sign at the Hill radius
\begin{equation} \label{eqn:hillradef}
  \rH \equiv \pfrac{m_c}{3 \Omega^2}^{1/3},
\end{equation}
separating the core-dominated potential from the tidally-dominated potential. The shock waves form near this transition, causing the steep increase in pressure support. The total radial acceleration is negligible inside $r<\rH$, so the envelope is indeed in steady state. The gravitational acceleration is balanced by inertial (Coriolis + centrifugal) and pressure support in roughly equal proportions, so the envelope deviates significantly from a hydrostatic solution. 

%%%%%%%%%%%%%%%%%%%%%%%%%%%%%%%%%%%%%%%%%%%%%%%%%%%

\subsection{Rotational support}

The departure from a hydrostatic equilibrium can be measured by comparing the density to its hydrostatic value. For isothermal flows with azimuthal symmetry, the hydrostatic density profile is
\begin{equation} \label{eqn:rhohsisot}
  \rho_{\mathrm{hs}}(r) = \rho_0 \,\exp\left[ - \frac{\Phi(r)}{c_s^2} \right],
\end{equation}
depending only on the \emph{local potential} $\Phi$ and sound speed $c_s$. In the opposite limit of no presure support, a flat density profile can be sustained by fully rotational support if the azimuthal velocity equals\footnote{We neglect the contribution of the Coriolis acceleration to the total inertial support; this term is always small near the core of rotationally supported envelopes, see \autoref{sec:circarcore}.} $\vK \equiv \sqrt{r \partial_r \Phi}$. 

\begin{figure}%[H]
\begin{center}
\includegraphics[width=1.0\columnwidth]{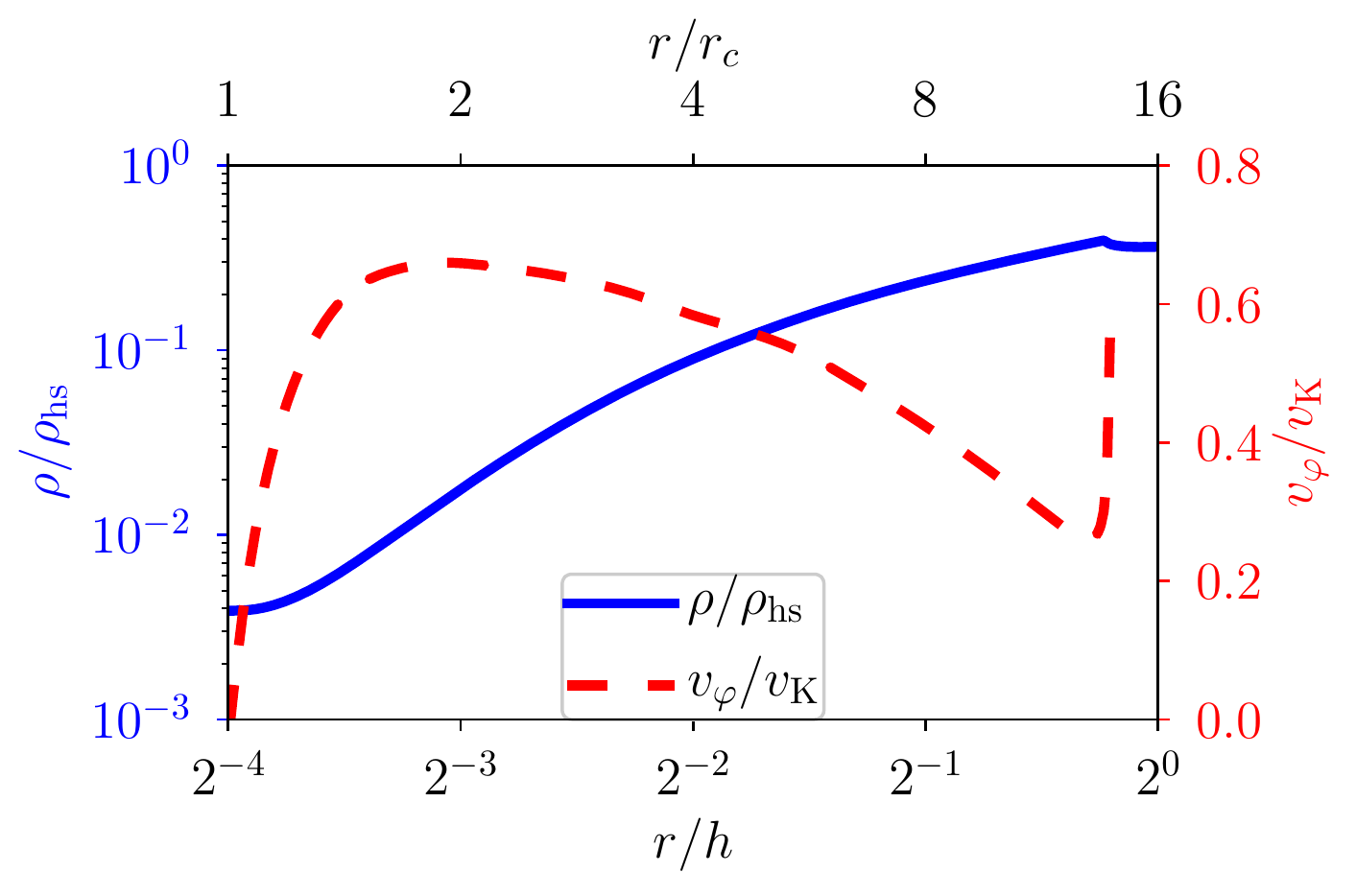}
\caption{Azimuthally-averaged profiles in the fiducial run \texttt{IH16B16}: density relative to the hydrostatic profile $\rho_{\mathrm{hs}}$ (solid blue, left axis) and angular velocity relative to fully rotational support $\vK$ (dashed red, right axis); note that there can be no rotational support beyond the Hill radius $\rH$ (which is not too different from $h$ in this run). \label{fig:i2dp16b16_SH_staticrhovp}}
\end{center}
\end{figure}

The density and angular velocity profiles of run \verb|IH16B16| are drawn on \autoref{fig:i2dp16b16_SH_staticrhovp}, respectively normalized by $\rho_{\mathrm{hs}}$ and $\vK$. The density always increases near the core surface, but not as fast as predicted by \eqref{eqn:rhohsisot}; the density at the core surface reaches only $4\times 10^{-3}$ of its hydrostatic value in this run. The importance of rotational support is signified by $v_{\varphi}/\vK$ reaching up to $70\%$ in the envelope. The azimuthally-averaged gravitational acceleration is outward outside approximately one Hill radius (see \autoref{fig:i2dp16b16_SH_axel}) because of Keplerian shear in the disk, so there can be no rotational support beyond. 

%%%%%%%%%%%%%%%%%%%%%%%%%%%%%%%%%%%%%%%%%%%%%%%%%%%

\subsection{Vortensity conservation and generation} 
\label{sec:vortensity}

Rotationally supported envelopes form in our simulations as a result of vortensity conservation. In inviscid barotropic flows, the vortensity $\varpi \equiv \left( \nabla\times v + 2\Omega \right) / \rho$ evolves according to
\begin{equation} \label{eqn:vortensinduc}
  \partial_t \varpi + v\cdot\nabla\varpi = \varpi\cdot\nabla v. 
\end{equation}
In two dimensions, the axial component $\varpi_z$ can vary only at shocks \citep[e.g.,][]{dong11}; it is otherwise conserved along streamlines. This quantity determines the amount of rotational support achieved in the envelope \citep[see][and \autoref{sec:circarcore}]{ormel1}.

\begin{figure}%[H]
\begin{center}
\includegraphics[width=\columnwidth]{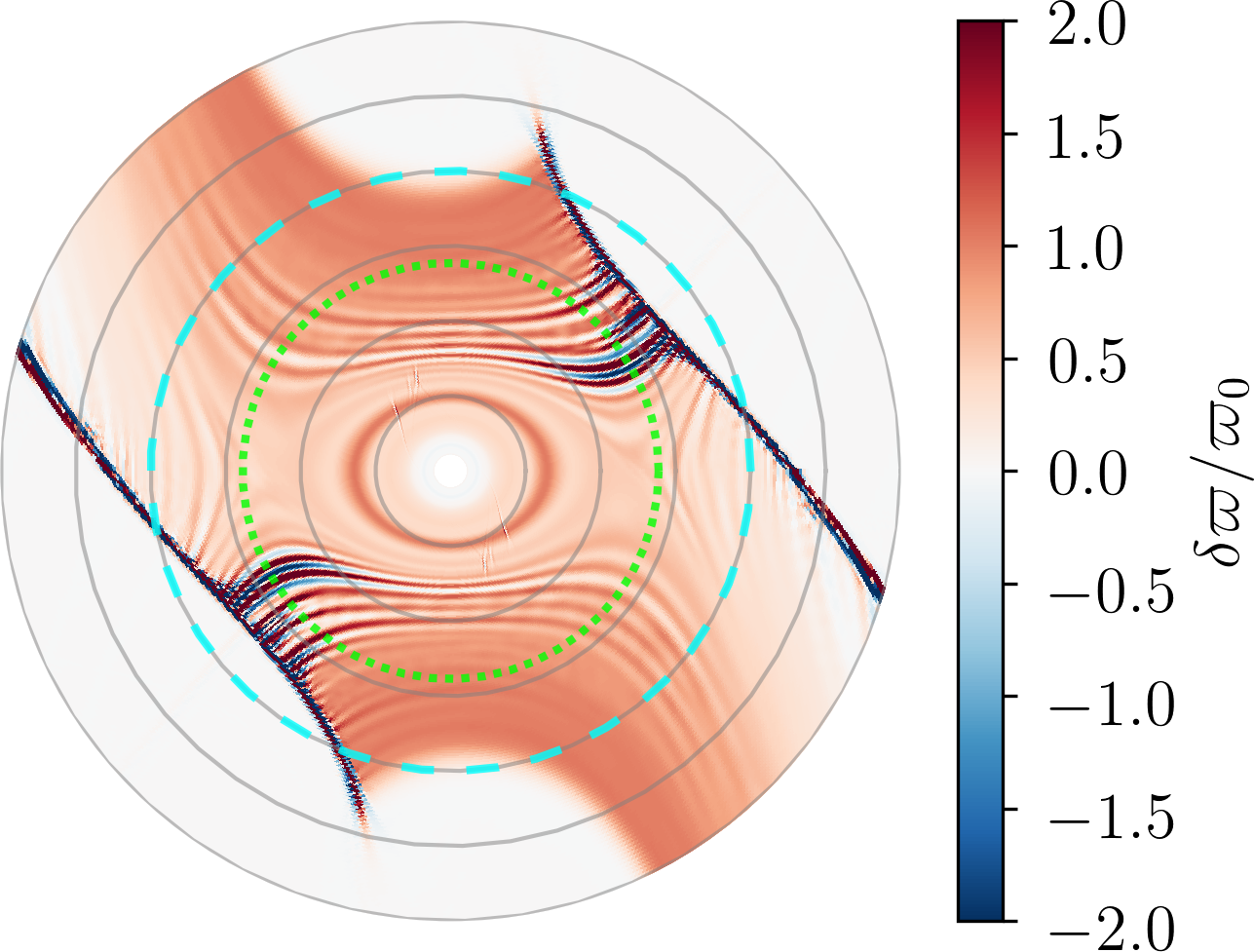}
\caption{Time-averaged vortensity fluctuations $(\varpi_z-\varpi_0)/\varpi_0$ in run \texttt{IH16B16}; the concentric circles mark every $4r_c$ (solid grey), the pressure scale $h$ (dashed-cyan) and the Hill radius $\rH$ (dotted green). The shock front, at which vortensity is generated, extends further away from the core compared to \autoref{fig:i2dp16b16_Map_rho}. \label{fig:i2dp16b16_Map_wda}}
\end{center}
\end{figure}

In the regime $B \gtrsim H$, the stationary shock waves visible in \autoref{fig:i2dp16b16_Map_rho} allow vortensity generation. \autoref{fig:i2dp16b16_Map_wda} reveals substantial deviations from the Keplerian vortensity value $\varpi_0$ in run \verb|IH16B16|. Vortensity is generated at the shocks and advected downstream. The shock fronts extend upstream of the core into the co-orbital region (lower left and upper right). This material gains a significant amount of vorticity despite the shocks being weak there ($\Delta \rho / \rho \rightarrow 0$, see \autoref{fig:i2dp16b16_Map_rho}). 

The flow obtained in this simulation is not strictly steady, and \autoref{fig:i2dp16b16_Map_wda} reveals filamentary patterns along the shock front in the post-shock medium. These filaments are only resolved by a few grid cells, but they do not vanish upon time averaging. They appear in simulations with $B \geq H$ and always develop along the shock front, with no clear signature in the density distribution. With a local shear rate comparable to the rotation rate of the flow, it is tempting to interpret these filamentary patterns as marginally stable sheared waves; a detailed stability analysis is however beyond the focus of this paper. No such features appear in the low-mass regime $B/H<1$, for which the flow is stationary in agreement with \cite{ormel1}.

%%%%%%%%%%%%%%%%%%%%%%%%%%%%%%%%%%%%%%%%%%%%%%%%%%%

\subsection{Mass and momentum transport}

In our simulations the flow features small-scale fluctuations over a quasi-steady state. This scale separation allows us to represent a flow variable $X$ as a laminar component $\overline{X}$ (moving average over time) plus short-timescale fluctuations $X'$. Under this decomposition, the radial mass transport obeys
\begin{equation} \label{eqn:turbmassflux}
  \partial_t\brac{\bar{\rho}} = - \frac{1}{r} \partial_r \left[ r \brac{\bar{\rho}\,\bar{v} + \overline{\rho' v_r'}}\right],
\end{equation}
where the second term in the right-hand side comes from correlations of density and velocity fluctuations. These two contributions to mass flux are respectively labeled as `laminar' and `turbulent'. We estimate the time-averages $\overline{X}$ by stacking $21$ snapshots spanning $10$ orbits in the quasi-steady state of run \verb|IH16B16|. We compute the fluctuations by subtracting $X'=X-\overline{X}$ in each snapshot; we then compute the correlations between fluctuating terms and average them over the same set of snapshots. 

\begin{figure}%[H]
\begin{center}
\includegraphics[width=\columnwidth]{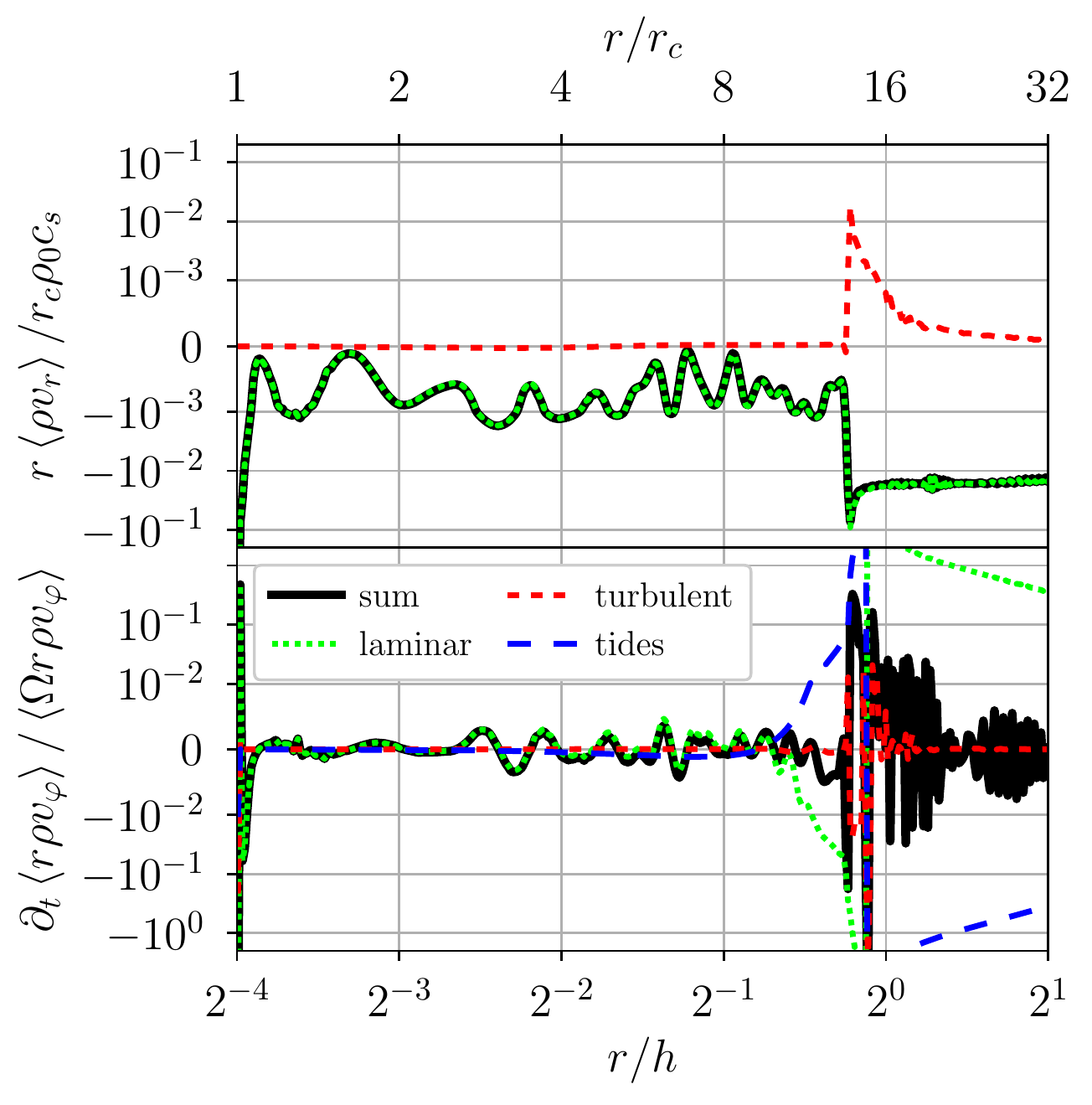}
\caption{Time- and azimuthally-averaged radial profiles in run \texttt{IH16B16}. \emph{Upper panel}: radial mass flux decomposed into laminar (dotted green) and turbulent (dashed red) components in the right-hand side of \eqref{eqn:turbmassflux}. Note that the mass flux at the inner boundary is counterbalanced from the ghost cells, so that no mass is lost from the computational domain into the core. \emph{Lower panel}: evolution of the angular momentum density due to gravitational torques $\rho \partial_{\varphi}\Phi$ (`tides', dashed blue) plus laminar and turbulent components of the momentum flux (remaining terms in \eqref{eqn:turbangmflux}). \label{fig:i2dp16b16_SH_masstorques}}
\end{center}
\end{figure}

The laminar and turbulent mass fluxes are drawn on the upper panel of \autoref{fig:i2dp16b16_SH_masstorques}. The turbulent mass flux is negligible inside the envelope. The laminar mass flux is oriented toward the core and causes mass accumulation near the core surface. The mass contained inside the Bondi radius increases by only $0.2\%$ over ten orbits, so the flow is quasi-steady to a high degree. This increment corresponds to the mass flux passing through the Bondi radius, there is no measurable mass loss into the core through the inner radial boundary. 

The same analysis can be performed for the radial transport of angular momentum, as seen in the non-inertial frame: 
\begin{equation} \label{eqn:turbangmflux}
  \partial_t \brac{r \rho v_{\varphi}} = - \brac{\rho \,\partial_{\varphi}\Phi + 2 \Omega r \rho v_r} - \frac{1}{r} \partial_r \brac{r \left(r \rho v_{\varphi} \right) v_r }.
\end{equation}
The first bracket includes gravitational tides and the Coriolis acceleration. The second bracket corresponds to the radial flux of momentum advected by the flow. We decompose this flux into one laminar and three turbulent components:
\begin{equation}
    \overline{\rho v_r v_{\varphi}} = \bar{\rho}\,\bar{v}_r\bar{v}_{\varphi} + \bar{\rho}\, \overline{v_r' v_{\varphi}'} + \overline{\rho' v_r'} \, \bar{v}_{\varphi} + \overline{\rho' v_{\varphi}'} \, \bar{v}_r.
\end{equation}
The different terms of \eqref{eqn:turbangmflux} are shown on the lower panel of \autoref{fig:i2dp16b16_SH_masstorques}. Inside the envelope, the derivative $\partial_t \brac{r\rho v_{\varphi}}$ fluctuates around zero with amplitudes small relative to $\brac{\Omega r \rho v_{\varphi}}$. The angular momentum of the envelope is thus conserved to a good approximation. The turbulent momentum flux is small at all radii. At the envelope boundary $r\approx h$, the gravitational torque induced by the star excites density waves that carry angular momentum away from the core. 

Since the net torque inside the envelope is negligible, mass accretion is driven by momentum losses at the envelope boundary. Numerical dissipation induces negligible accretion rates at the current resolution (cf. Appendix \ref{app:convstudy}), so the isothermal shocks are responsible for momentum dissipation. This accretion mechanism differs from viscously-driven accretion \citep[e.g.,][]{kley99} and primarily concerns massive cores in 2D \citep[see][and \autoref{sec:onseturb2d}]{lubow99}. 

%%%%%%%%%%%%%%%%%%%%%%%%%%%%%%%%%%%%%%%%%%%%%%%%%%%

\subsection{Envelope recycling}
\label{sec:recycle}

One of the goals of this study is to characterize the recycling of the envelope material for different core masses. As recycling represents the exchange of mass between the envelope near the core and the adjacent parts of the disc, unbound to the core, we need to come up with ways of characterizing this exchange. Here we describe two such methods, utilizing a passive tracer \autoref{sec:fidpasstracfluid} and gravitational binding of the envelope \autoref{sec:fidgravbound}, and illustrate their application in our fiducial run.

%%%%%%%%%%%%%%%%%%%%%%%%%%%%%%%%%%%%%%%%%%%%%%%%%%%

\subsubsection{Tracer fluid} 
\label{sec:fidpasstracfluid}

To examine the mixing properties of the flow, we follow the transport of a passive tracer in run \verb|IH16B16|. We let the flow evolve toward a quasi-steady state over eight orbits after the core potential has been fully introduced; we then inject a tracer with concentration $n=1$ inside the Bondi disk. The tracer concentration is then advected by the flow, obeying
\begin{equation}
(\partial_t+v\cdot\nabla)n = 0. 
\end{equation}
We set the tracer concentration to zero in the radial ghost cells, so the boundaries can only absorb the traced fluid. 

\begin{figure}%[H] 
\begin{center}
\includegraphics[width=1.0\columnwidth]{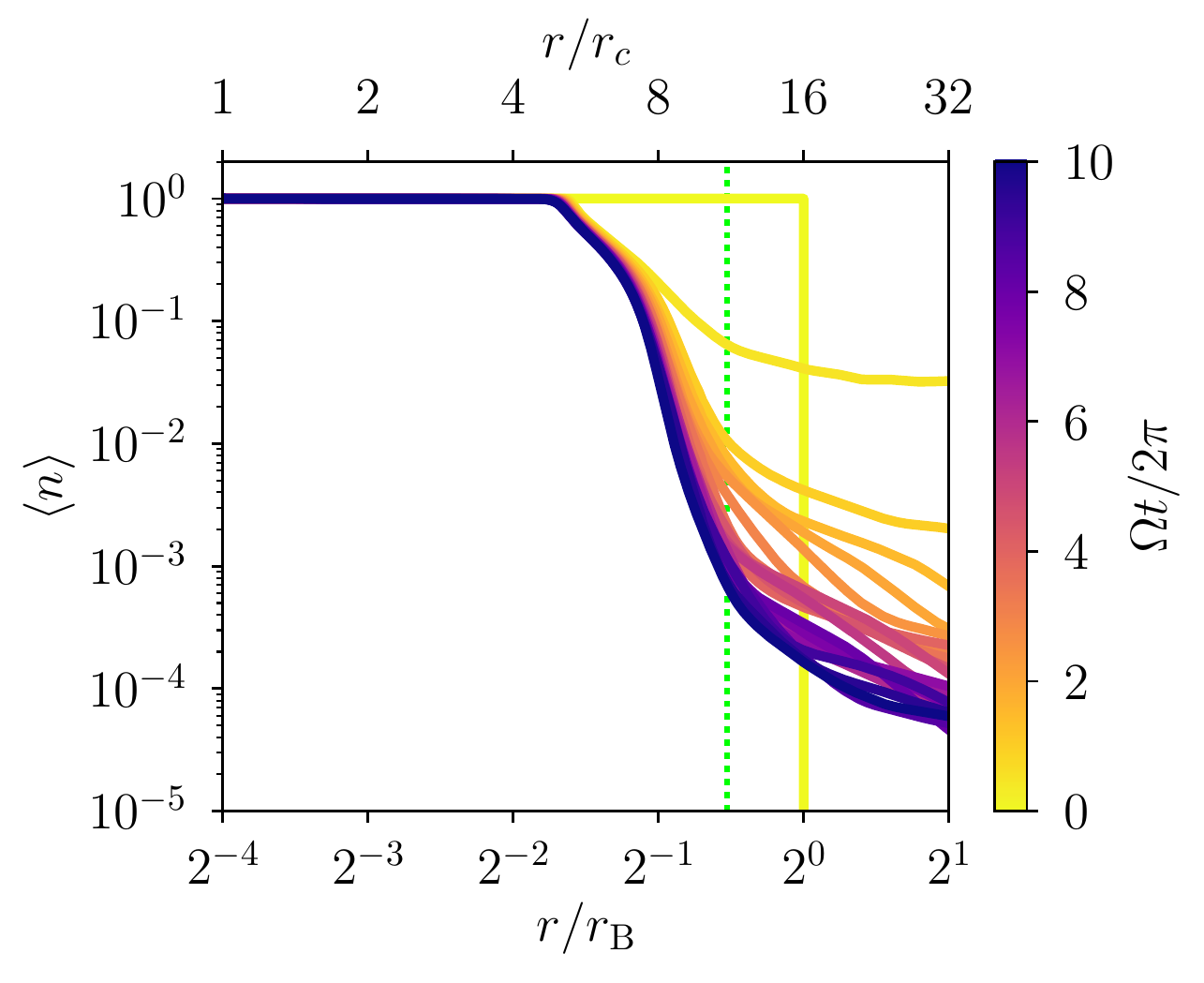}
\caption{Tracer concentration over $10$ orbits since injection in run \texttt{IH16B16}; the plateau $\langle n\rangle =1$ at small radii corresponds to the region of closed streamlines around to the core. The vertical dotted line marks the Hill radius. \label{fig:i2dp16b16_SH_tr1n}}
\end{center}
\end{figure}

\autoref{fig:i2dp16b16_SH_tr1n} shows the evolution of the tracer concentration over $10$ orbits since injection in run \verb|IH16B16|. At small radii $r/\rB \lesssim 1/2$, the tracer concentration remains unity, meaning that this region does not exchange matter with the outer envelope. This region is defined by having only closed streamlines in the time-averaged (`laminar') flow, confirming that turbulent velocity fluctuations do not contribute to gas mixing in the inner envelope. The tracer concentration drops by more than three orders of magnitude near $r/\rB \approx 1/2$, beyond which it is advected away from the core in less than two orbits. Recycling thus appears to be either switched on or off, depending on the topology of the time-averaged flow alone. In principle, the fluid orbiting on closed streamlines around the core could radiate its internal energy, and eventually remain bound to the core by gravity. To determine the extent of such a gravitationally bound envelope, we move on to an energetic arguments below.

%%%%%%%%%%%%%%%%%%%%%%%%%%%%%%%%%%%%%%%%%%%%%%%%%%%

\subsubsection{Gravitationally bound envelopes} \label{sec:fidgravbound}

If the nebula in which the core is embedded were to disperse with time, the envelope would expand to match the reduced ambient pressure \citep{owenjackson12}. For the core to keep a gravitationally bound atmosphere, the energetic content of the gas must not allow it to escape the potential well of the core. The critical energy barrier $\Phi_{\mathrm{H}}$ is the total potential \eqref{eqn:totpotgrav} evaluated at one Hill radius. Whether the envelope is gravitationally bound can be assessed via the Bernoulli number
\begin{equation} \label{eqn:bernoullidef}
  \mathcal{B} \equiv \frac{v^2}{2} + \mathcal{H} + \Phi - \Phi_{\mathrm{H}}, 
\end{equation}
where the enthalpy is $\mathcal{H} \equiv c_s^2 \log \rho$ in isothermal flows. We verified that $\mathcal{B}$ is accurately conserved along streamlines in the time-averaged flow of our simulations. A gravitationally bound streamline should therefore satisfy $\mathcal{B} < 0$. 

\begin{figure}%[H]
\begin{center}
\includegraphics[width=\columnwidth]{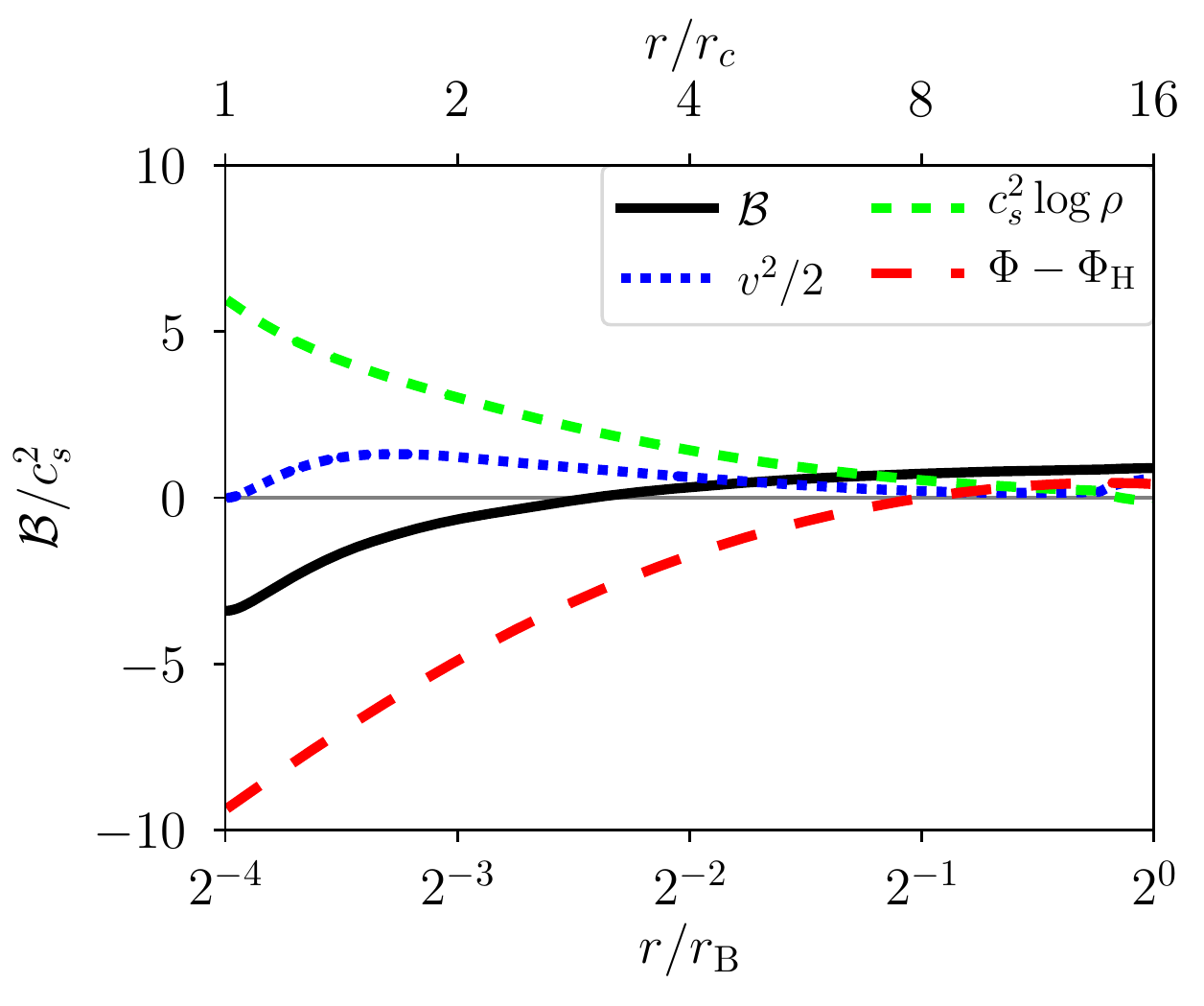}
\caption{Bernoulli number in run \texttt{IH16B16}, split into its kinetic ($v^2/2$, dotted blue), internal (enthalpy, dashed green) and gravitational ($\Phi$, dashed red) contributions. \label{fig:i2dp16b16_SH_brnl}}
\end{center}
\end{figure}

\autoref{fig:i2dp16b16_SH_brnl} represents the radial distribution of Bernoulli number and its different contributions in run \verb|IH16B16|. The kinetic term is smaller than the enthalpy contribution despite the flow orbiting the core at sonic velocity $v_{\varphi}/c_s \approx 1$. The Bernoulli number becomes negative inside $r/\rB \lesssim 1/5$, corresponding to approximately $3 r_c$ in this case. Since the flow is immediately maintained to its initial temperature, neither adiabatic compression nor shocks can heat the envelope. Isothermal envelopes represent the coldest solutions given a background disk temperature, so they provide an upper bound on the size of the gravitationally bound region. As long as the kinetic contribution is small, the Bernoulli number is only a function of $\rB/r$ inside the Hill radius; the radius at which $\mathcal{B}=0$ should therefore be a fraction of the Bondi radius. We consistently find this radius near $\rB/5$ in every isothermal simulation, regardless of $B/H$. If the disk were to disperse, we would expect the outer parts of the envelope $r/\rB \gtrsim 1/5$ to expand and escape the gravity of the core. This region includes closed streamlines, unaffected by recycling under the current pressure confinement from the disk onto the envelope.

%%%%%%%%%%%%%%%%%%%%%%%%%%%%%%%%%%%%%%%%%%%%%%%%%%%

\section{Parameter exploration: isothermal simulations} 
\label{sec:isotsimu}

%%%%%%%%%%%%%%%%%%%%%%%%%%%%%%%%%%%%%%%%%%%%%%%%%%%

We now examine how the outcomes found in a fiducial case change as we vary $B$ and $H$ in a series of isothermal simulations. This corresponds to changing the mass of the core $m_c = B\,H^2$ (in units of $m_{\rm th}$) and its radius relative to the disk scale height $r_c/h = H^{-1}$. 

%%%%%%%%%%%%%%%%%%%%%%%%%%%%%%%%%%%%%%%%%%%%%%%%%%%

\subsection{Pressure versus rotational support}

\begin{figure}%[H]
\begin{center}
\includegraphics[width=\columnwidth]{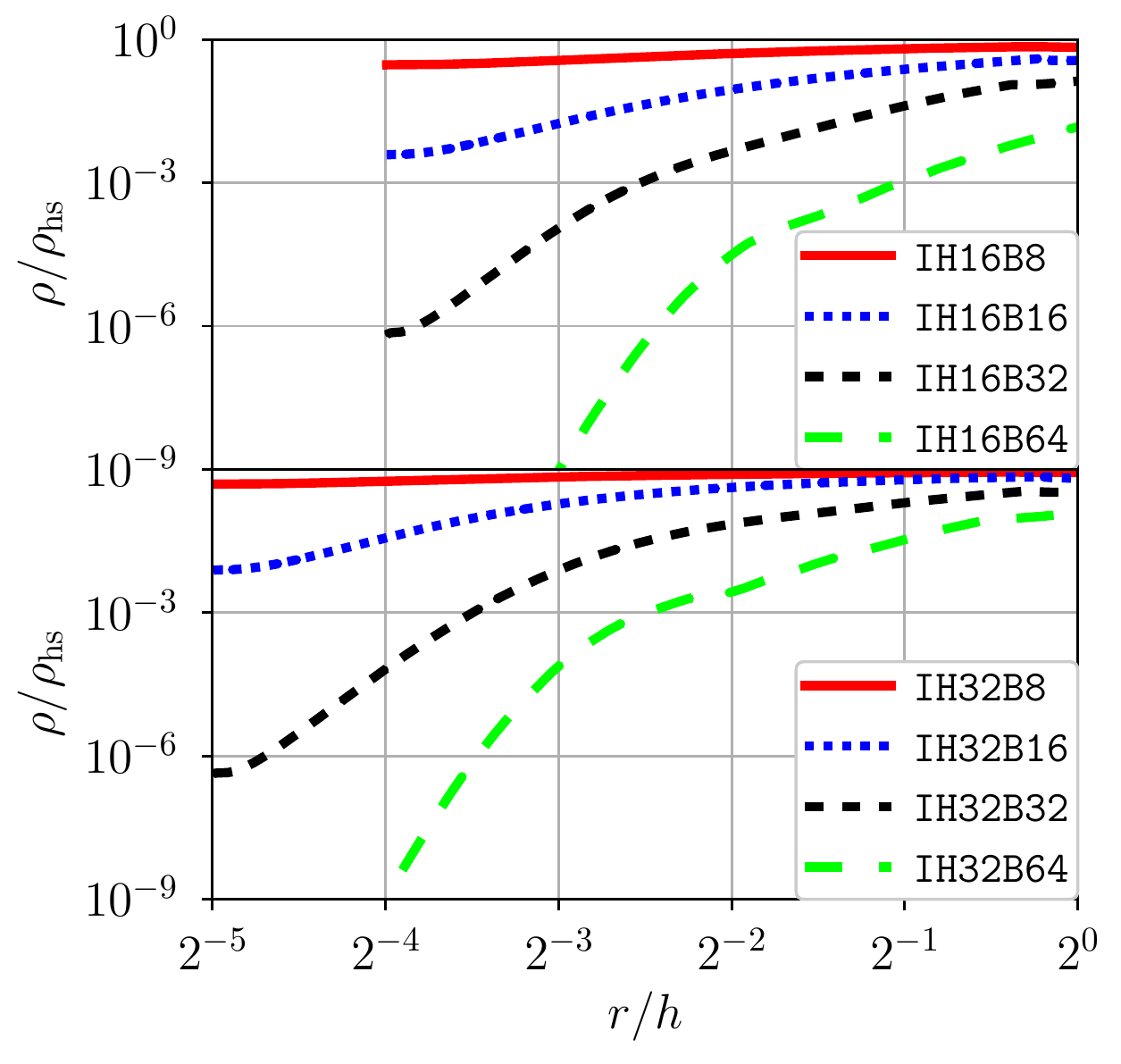}
\caption{Azimuthally-averaged density normalized to the hydrostatic density \eqref{eqn:rhohsisot} for $H=16$ (upper panel) and $H=32$ (lower panel) and for different values of $B$ (see legend). \label{fig:i2dp16-32_rhohs}}
\end{center}
\end{figure}

\autoref{fig:i2dp16-32_rhohs} displays the radial profiles of density relative to the hydrostatic value $\rho_{\rm hs}$ in a series of isothermal simulations. For a fixed $H$ (upper or lower panel), increasing $B$ (the planet mass) leads to progressively higher densities at the same radius. However, the density does not increase as fast as predicted for a hydrostatic profile \eqref{eqn:rhohsisot}; the ratio $\rho / \rho_{\mathrm{hs}}$ decreases with $B$, as \autoref{fig:i2dp16-32_rhohs} shows. Comparing the two panels, we see that curves with the same $B$ but different $H$ are essentially identical but offset by a factor $2$ horizontally. The amount of pressure support $\rho / \rho_{\mathrm{hs}}$ thus depends on $B$ and on the distance to the core surface, but only weakly on $H$. In other words, the ratio of $B/H$, independent of $r_c$, does not uniquely determine the amount of pressure support as a function of $r/h$. 

As $\rho / \rho_{\mathrm{hs}}$ decreases, the amount of rotational support $v_{\varphi}/\vK$ increases toward unity. Therefore, the amount of rotational support in the entire envelope also depends on $r_c/h$. We show in \autoref{sec:circarcore} that this dependence on the core radius is expected to vanish in the limit $H \rightarrow \infty$. 

% HERE

%%%%%%%%%%%%%%%%%%%%%%%%%%%%%%%%%%%%%%%%%%%%%%%%%%%

\subsection{Tracer fluid} 

\begin{figure}%[H]
\begin{center}
\includegraphics[width=\columnwidth]{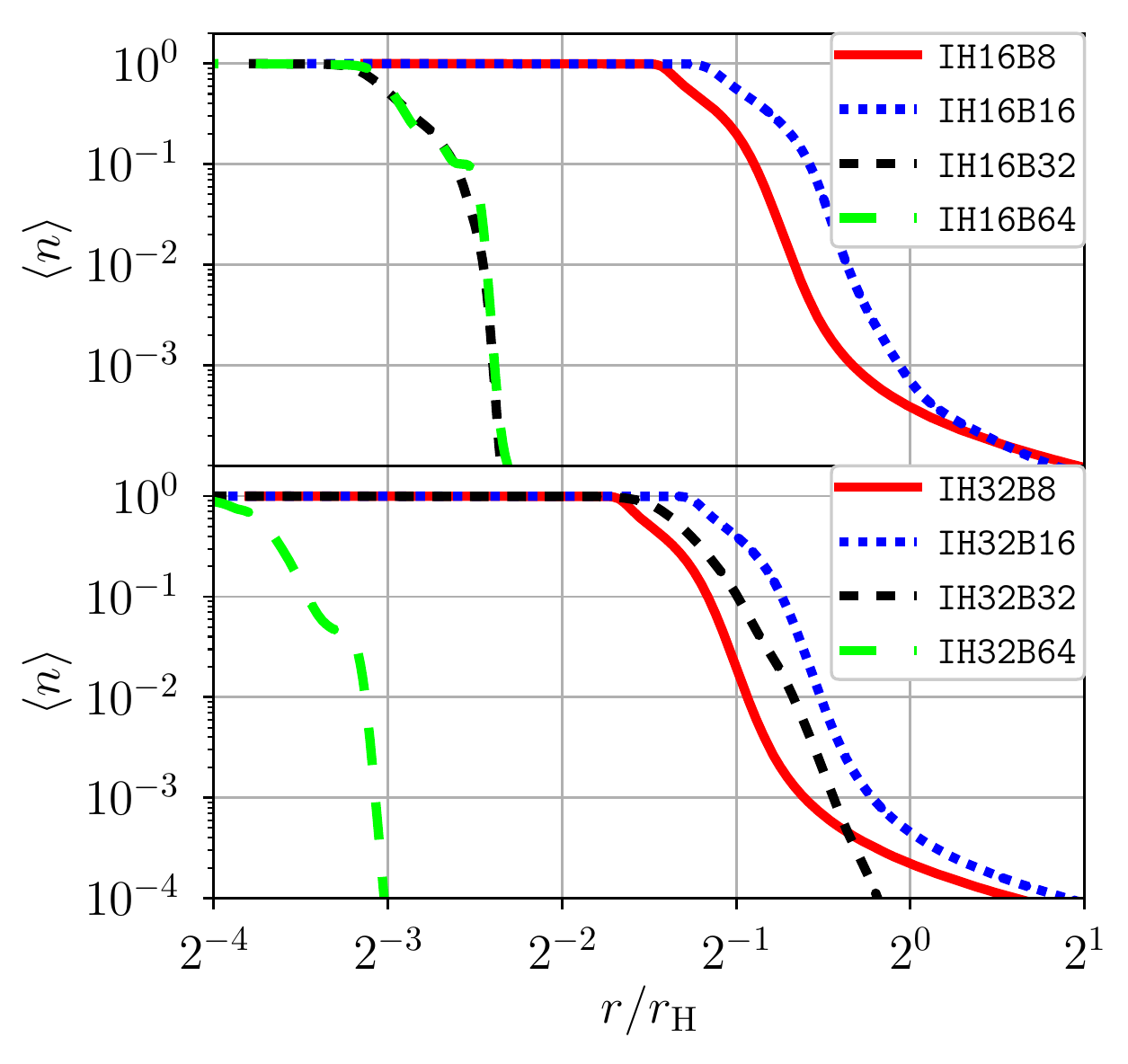}
\caption{Azimuthally-averaged tracer concentration after ten orbits in isothermal runs with $H=16$ (upper panel), $H=32$ (lower panel), and different values of $B$ (see legend). \label{fig:i2d_tracer}}
\end{center}
\end{figure}

After the potential of the core is introduced, we let the flow settle to a quasi-steady state over eight orbits in simulations with different values of $H$ and $B$. We then inject a passive tracer fluid with concentration $n=1$ inside the Bondi disk, and let it evolve over ten orbits. The final tracer distributions are drawn on \autoref{fig:i2d_tracer}. 

As in \autoref{sec:fidpasstracfluid}, every simulation features a concentration plateau below a critical radius, so tracer mixing is inefficient in the direct vicinity of the core. Comparing both panels one notes that, for a fixed value of $H$, this critical radius remains near $\rH/3$ as long as $B\leq H$, i.e. $m_c\leq m_{\rm th}$. For larger values of $B>H$ the critical radius moves much closer to the core, meaning that most of the envelope was recycled by the disk flow over the duration of the simulation. Closed streamlines are still present in the time-averaged flow of every simulation, but a fraction of these streamlines have a reduced tracer concentration. We therefore attribute recycling to time-dependent fluctuations. \autoref{fig:i2d_tracer} suggests the onset of turbulent tracer transport beyond a critical core mass $B/H>1$. 

% HERE

%%%%%%%%%%%%%%%%%%%%%%%%%%%%%%%%%%%%%%%%%%%%%%%%%%%

\subsection{Mass and tracer transport around massive cores} \label{sec:onseturb2d}

\begin{figure}%[H]
\begin{center}
\includegraphics[width=\columnwidth]{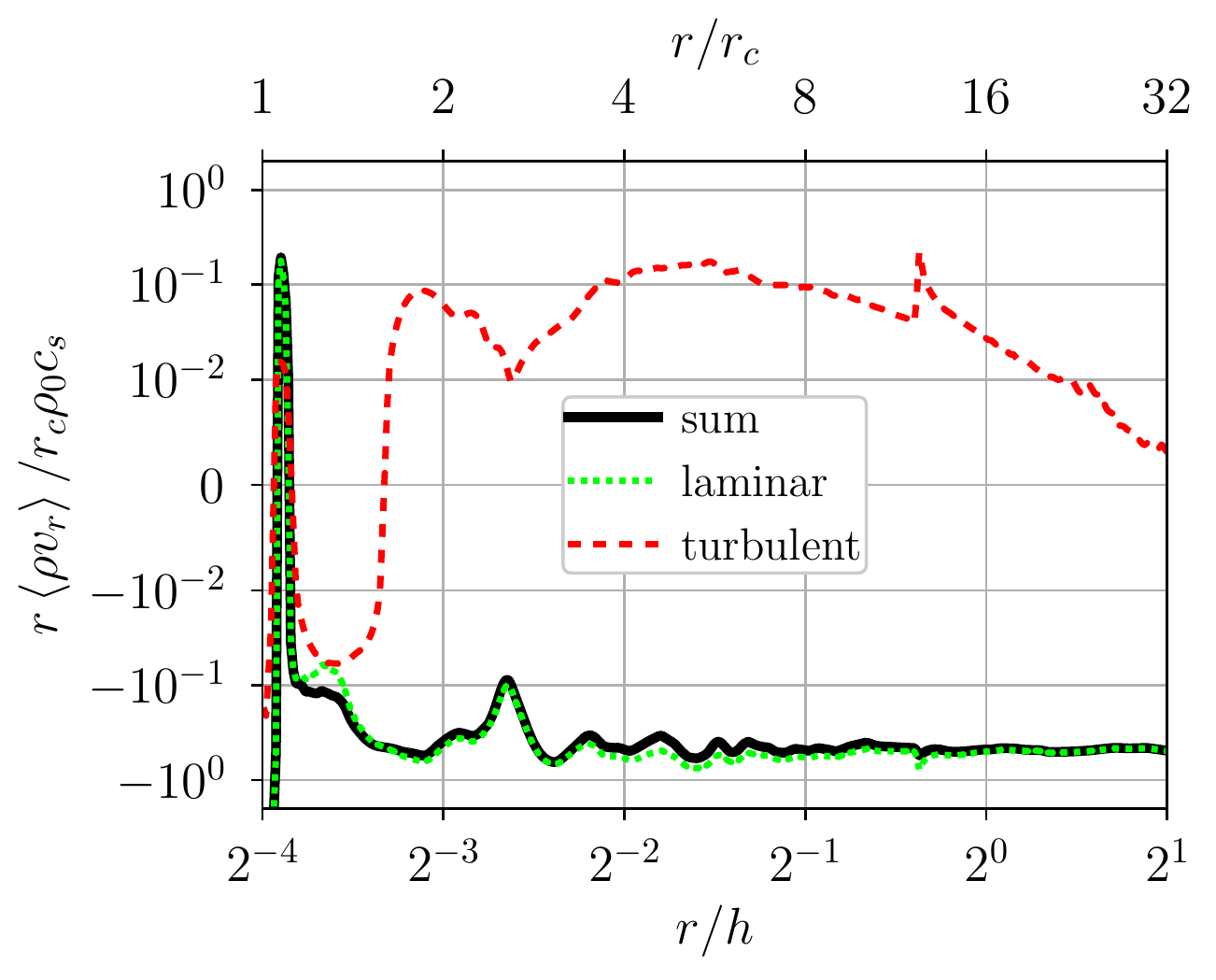}
\caption{Average radial mass flux in run \texttt{IH16B32} (solid black), split into laminar (dotted green) and turbulent (dashed red) components as in \eqref{eqn:turbmassflux}; compare with the fiducial (non-turbulent) case on \autoref{fig:i2dp16b16_SH_masstorques}. \label{fig:i2dp16b32_SH_massflux}}
\end{center}
\end{figure}

We recall that in run \verb|IH16B16| the envelope displays an extended tracer plateau (\autoref{fig:i2d_tracer}) and essentially no turbulent mass flux (\autoref{fig:i2dp16b16_SH_masstorques}). However, with a core twice more massive, run \verb|IH16B32| features tracer mixing much deeper in the envelope. We examine the radial mass fluxes in run \verb|IH16B32| on \autoref{fig:i2dp16b32_SH_massflux}.

The net mass flux is oriented towards the core, with an amplitude $\sim 10^3$ larger than in the fiducial case \texttt{IH16B16}. The mass contained within the Bondi disk increases by $40\%$ over ten orbits, corresponding to the mass flux passing through the Bondi radius. The mass flux through the inner radial boundary fluctuates around zero with amplitudes at the level of $\pm 10^{-6}m_c$ per orbit, negligible compared to mass accretion rate of the envelope. Compared to run \verb|IH16B16|, there is now a significant turbulent mass flux inside the envelope, reaching amplitudes larger than $10\%$ of the laminar one. The turbulent mass flux of \eqref{eqn:turbmassflux} allows mass to flow across the time-averaged velocity field. It allows the mixing of the envelope with the background flow, regardless of momentum transport.

\begin{figure}%[H]
\begin{center}
\includegraphics[width=1.0\columnwidth]{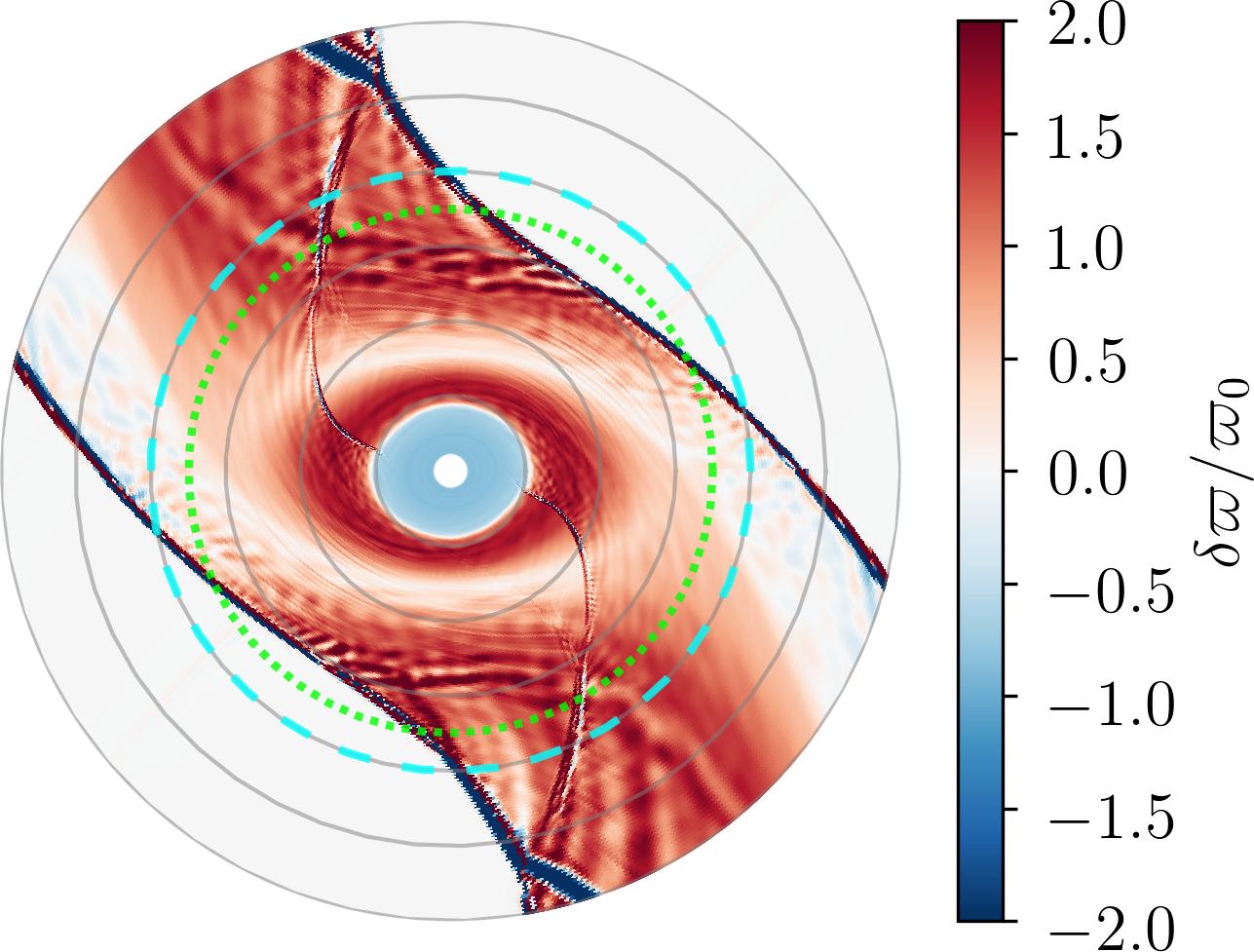}
\caption{Time-averaged vortensity fluctuations $(\varpi_z-\varpi_0)/\varpi_0$ in run \texttt{IH16B32}; the circles mark every $4r_c$ (solid grey), the pressure scale $h$ (dashed cyan) and the Hill radius $\rH$ (dotted green). Shocks in the rotating envelope produce a ring of increased vortensity near the core; compare with the fiducial case on \autoref{fig:i2dp16b16_Map_wda}. \label{fig:i2dp16b32_Map_wda}}
\end{center}
\end{figure}

\autoref{fig:i2dp16b32_Map_wda} shows the time-averaged vortensity distribution in run \verb|IH16B32|. Compared with \verb|IH16B16| on \autoref{fig:i2dp16b16_Map_wda}, we see that stationary shocks emerge inside the Hill radius. These shocks alter the vortensity distribution in the rotating envelope, the sign of the vortensity jump being sensitive to the geometry of the shock \citep{kevlahan97}. The core is surrounded by a low-vortensity disk extending to $\lesssim 4 r_c$, and delimited by a high-vortensity ring near $5 r_c$. The low-vortensity region encloses the remaining tracer fluid visible in \autoref{fig:i2d_tracer}. The shock locations are also correlated with the presence of a turbulent mass flux in \autoref{fig:i2dp16b32_SH_massflux}. We therefore attribute the envelope recycling to the formation of steady shocks in the circulating flow within the Hill sphere of the core.

Momentum dissipation at these shocks also drives the laminar mass flux toward the core. As vortensity decreases near the core, the role of rotational support decreases \citep[see][and \autoref{sec:circarcore}]{ormel1}. To maintain radial momentum balance, the envelope must increase its pressure support by accreting mass. Without including the gas gravity, this accretion process cannot produce envelopes more massive than the hydrostatic limit given by \autoref{eqn:rhohsisot}. On long time scales, the asymptotic distribution of vortensity (rotational support) will be sensitive to the viscous processes at play (e.g., turbulence) and cannot be uniquely determined a priori in an inviscid context.

%%%%%%%%%%%%%%%%%%%%%%%%%%%%%%%%%%%%%%%%%%%%%%%%%%%

\section{Parameter exploration: non-isothermal simulations} 
\label{sec:adiabatic}

%%%%%%%%%%%%%%%%%%%%%%%%%%%%%%%%%%%%%%%%%%%%%%%%%%%

Having explored the characteristics of the flow in isothermal setup, we now change our thermodynamic assumptions and simulate the flow with a adiabatic equation of state $P \propto \rho^{\gamma}$. The energy equation \eqref{eqn:conse} is integrated conservatively, and the entropy per unit mass $s$ is conserved along streamlines as long as the flow does not shock. This situation can be relevant in the optically thick limit, when the heat due to adiabatic compression is not efficiently radiated away. The exponent is $\gamma = 7/5$ everywhere and at all time, neglecting chemical and ionization effects. Fresh gas with the entropy of the background disk is continually supplied at the outer radial boundary. We still expect convergence to quasi-steady states, as no energy losses are allowed and entropy production at the shocks is rather slow. The control parameters $H$ and $B$ (which depend on temperature) now represent the thermodynamic properties of the background disk flow. 

%%%%%%%%%%%%%%%%%%%%%%%%%%%%%%%%%%%%%%%%%%%%%%%%%%%

\subsection{Adiabatic versus irreversible heating} 
\label{sec:thermalstruct}

%%%%%%%%%%%%%%%%%%%%%%%%%%%%%%%%%%%%%%%%%%%%%%%%%%%

As the core mass is progressively introduced in our adiabatic runs, the envelope contracts and heats adiabatically, the temperature of the gas $T\equiv P/\rho$ increasing as $\rho^{\gamma-1}$. Once the gravitational potential of the core is set, the gas still heats and cools adiabatically as it passes by the core. In addition to adiabatic heating, the shocks forming at the envelope boundary induce irreversible heating, changing entropy $s$ of the gas \citep{rafikov16}. To measure the amount of irreversible heating at shocks, we compute the potential temperature
\begin{equation} 
\label{eqn:potmpdef}
  \Theta \equiv T \pfrac{P_{0}}{P}^{\left(\gamma-1\right)/\gamma} \sim \exp\left(s\right). 
\end{equation}
This quantity represents the temperature that a fluid element would have if its pressure were adiabatically relaxed to the background disk pressure $P_{0}$. Being only a function of the local entropy, $\Theta(s)$ is also conserved along streamlines and increases at shocks. We normalize it to the background disk temperature $T_{0} = H^2$. 

\begin{figure}%[H]
\begin{center}
\includegraphics[width=\columnwidth]{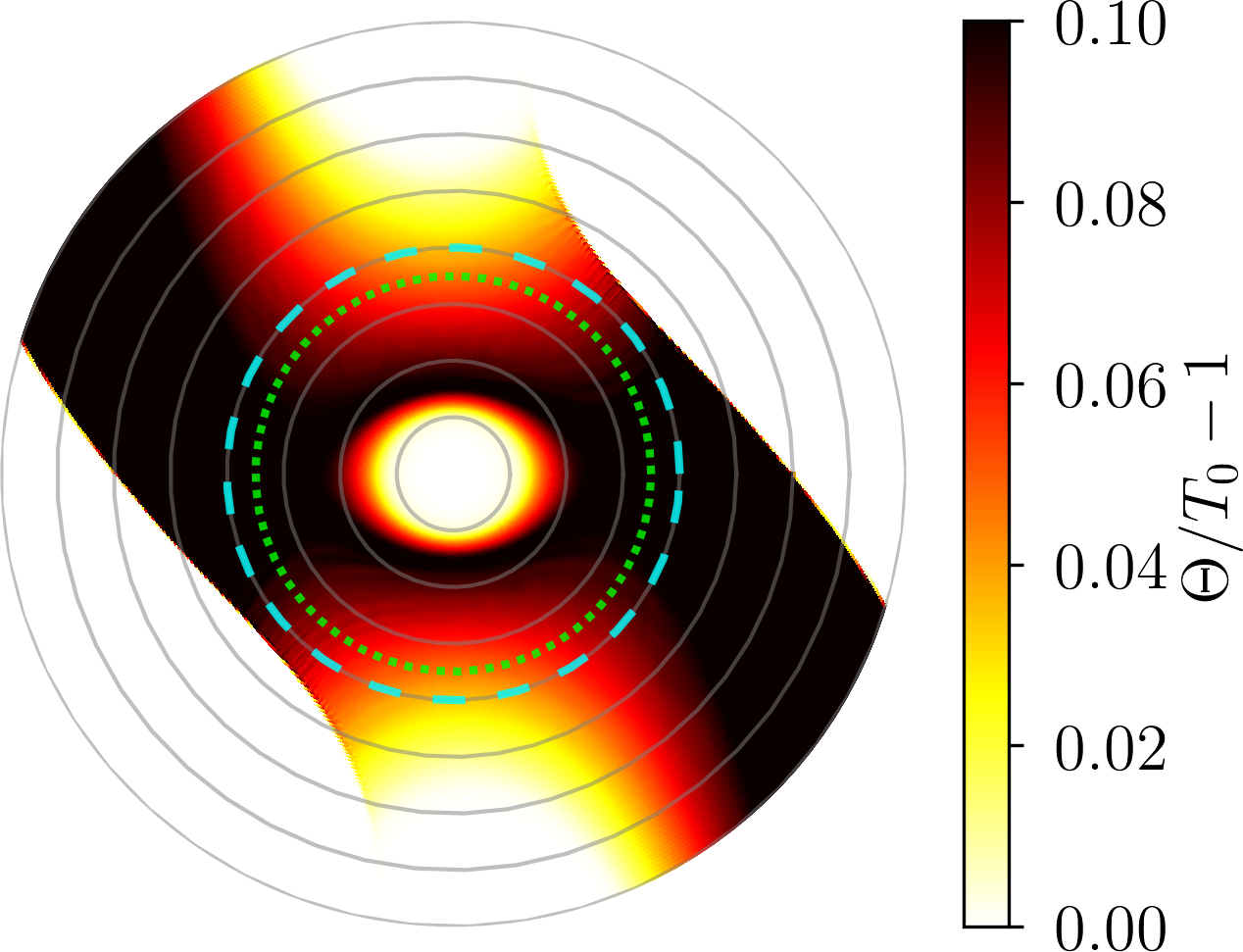}
\caption{Time-averaged deviations of potential temperature $\Theta$ relative to the background disk temperature $T_{0}$ in run \texttt{AH16B32}; the circles mark every $4r_c$ (solid grey), the pressure scale $h$ (dashed cyan) and the Hill radius $\rH$ (dotted green). Shocks induce irreversible heating at the envelope boundary, but the shocked flow does not reach the vicinity of the core. \label{fig:a2dp16b32_Map_potmp}}
\end{center}
\end{figure}

\autoref{fig:a2dp16b32_Map_potmp} shows the distribution of potential temperature in the quasi-steady state of run \verb|AH16B32|. Compared to the isothermal equivalent (see \autoref{fig:i2dp16b32_Map_wda}), the incoming gas shocks further away from the core in the adiabatic case, so the envelope spans a larger area. The potential temperature increases by up to $10\%$ at the envelope boundary; it is then passively advected inside and out of the envelope. The absence of irreversible heating near the core implies that the shocked gas never blends into the inner envelope. 

\begin{figure}%[H]
\begin{center}
\includegraphics[width=\columnwidth]{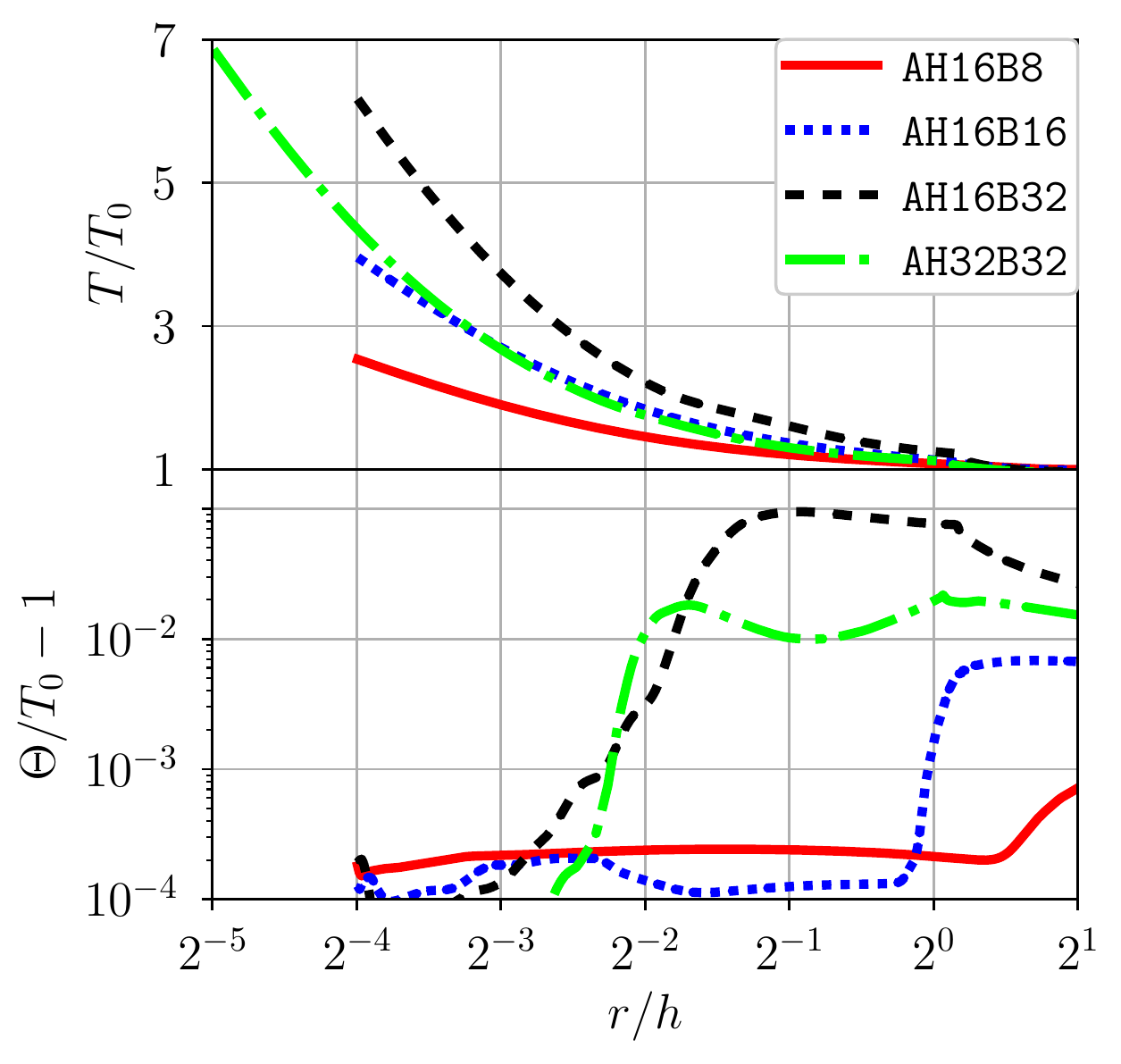}
\caption{Azimuthally-averaged profiles of temperature $T$ (upper panel) and deviations of the potential temperature $\Theta$ (lower panel) relative to the background temperature $T_{0}$ in four adiabatic simulations (see legend). 
\label{fig:a2d_tmpptm}}
\end{center}
\end{figure}

The upper panel of \autoref{fig:a2d_tmpptm} shows the radial profiles of temperature in a series of adiabatic simulations. The temperature always decreases with radius and increases with the mass of the core. The temperature at the surface of the core is up to seven times larger than its background value in run \texttt{AH32B32}. The temperature profiles of runs \texttt{AH16B16} and \texttt{AH32B32} are nearly superimposed as a function of $r/h$. Unlike isothermal envelopes, the radial structure of adiabatic envelopes apparently depends only on $B/H$ and not on the core radius. This behavior is expected if rotational support is small and if the entropy remains close to its background value \citep[e.g., equation 52 of][]{rafikov06}.

The lower panel of \autoref{fig:a2d_tmpptm} shows the radial profiles of potential temperature in the same series of adiabatic simulations. We observe irreversible heating for $B\gtrsim H$, but the potential temperature variations remain below $10\%$ for the range of parameters considered. The shock-induced heating is small compared to the total heating (see the upper panel of \autoref{fig:a2d_tmpptm}), so these envelopes can be considered as isentropic to a good approximation. In runs \texttt{AH16B8} and \texttt{AH16B16}, the spiral shocks do not extend upstream of the core into the co-orbital flow, as is the case on \autoref{fig:a2dp16b32_Map_potmp}. The shocked material is advected away without mixing into the envelope, so the heat deposition is localized near $r\approx h$. In runs \texttt{AH16B32} and \texttt{AH32B32}, the shocked gas blends into the envelope and mildly increases its entropy down to $r \gtrsim h/8$. Inside $r\lesssim h/8$, the potential temperature keeps its initial value, implying that the shocked gas never enters the innermost regions of these envelopes. 

%%%%%%%%%%%%%%%%%%%%%%%%%%%%%%%%%%%%%%%%%%%%%%%%%%%

\subsection{Envelope recycling} 
\label{sec:adiarecycle}

\begin{figure}%[H]
\begin{center}
\includegraphics[width=1.0\columnwidth]{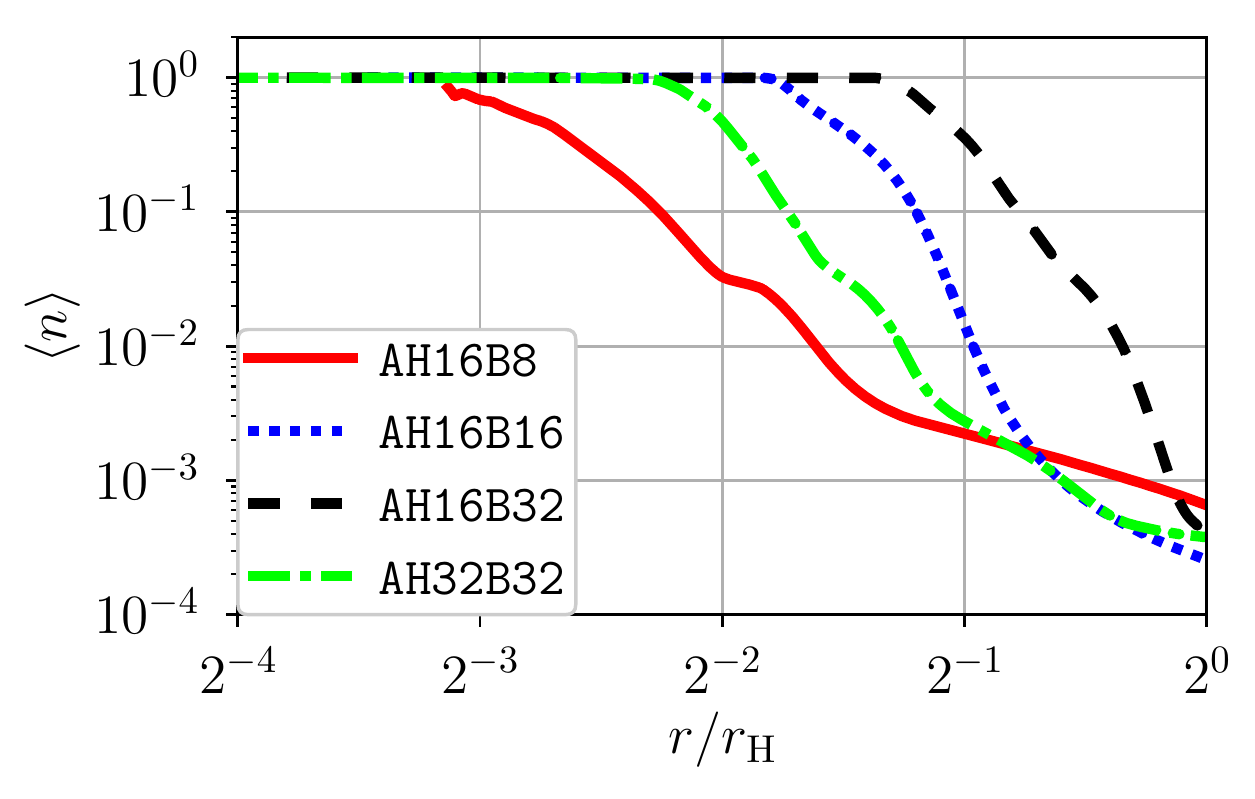}
\caption{Tracer concentration ten orbits after injection of tracer fluid inside the Bondi radius in adiabatic simulations; compare with the isothermal equivalent on \autoref{fig:i2d_tracer}. \label{fig:a2d_tracer}}
\end{center}
\end{figure}

To examine the efficiency of envelope recycling, we inject a tracer fluid in the quasi-steady state of each adiabatic simulation. The tracer concentration $n=1$ inside the Bondi disk and zero outside, and it is passively advected over ten orbits. The final concentration profiles are drawn on \autoref{fig:a2d_tracer}. Recycling affects the envelope inside the Hill radius in every case. Every run except \texttt{AH16B8} features a concentration plateau near the core. The radial extent of this plateau shows no obvious scaling with $B$ or $H$. Both runs \texttt{AH16B16} and \texttt{AH32B32} feature a concentration plateau extending to $\approx \rH/4$. Their concentration profiles are also similar to the isothermal equivalent \texttt{IH16B16} (see \autoref{fig:i2d_tracer}). Contrarily to their isothermal analogues, there is no concentration plateau near the core in run \texttt{AH16B8}, and there is an extended plateau in run \texttt{AH16B32}. 

In run \texttt{AH16B8}, the envelope is rotationally supported at only $v_{\varphi}/\vK \leq 3\%$, instead of $30\%$ in the equivalent isothermal run \texttt{IH16B8}. Only the innermost streamlines circulate around the core, so most of the envelope is recycled by the shear flow on orbital timescales. In run \texttt{AH16B32}, the envelope is rotationally supported at $v_{\varphi}/\vK\approx 30\%$, instead of $70\%$ in run \texttt{IH16B32}. Because of the temperature increase, the circulating flow remains subsonic in run \texttt{AH16B32}, with $v_{\varphi}/c_s \lesssim 60\%$ instead of $v_{\varphi}/c_s \in \left[1,3\right]$ in run \texttt{IH16B32}. We showed that stationary shocks develop in the circulating flow of run \texttt{IH16B32} (see \autoref{fig:i2dp16b32_Map_wda}), and we correlated these shocks to turbulent mixing (see \autoref{fig:i2dp16b32_SH_massflux}). Because the envelope of run \texttt{AH16B32} remains subsonic, no shocks develop and turbulent tracer transport does not operate. Laminar and circular streamlines extend up to $r\lesssim \rH/3$, corresponding to the extent of the tracer plateau visible on \autoref{fig:a2d_tracer}. 

%%%%%%%%%%%%%%%%%%%%%%%%%%%%%%%%%%%%%%%%%%%%%%%%%%%

\section{Parameter exploration: self-gravitating disks} 
\label{sec:selfgravity}

%%%%%%%%%%%%%%%%%%%%%%%%%%%%%%%%%%%%%%%%%%%%%%%%%%%

As the gas density near the core increases, the mass of the envelope could become comparable to the mass of the core. The gravitational acceleration induced by the gas could then alter the structure of the envelope even when background disk is not susceptible to a gravitational instability. We test the sensitivity of our previous results to inclusion of the gas self-gravity in a series of isothermal simulations. 

Gravitational instability is expected to occur in massive disks, when the Toomre parameter 
\begin{equation}
  \begin{aligned}
    Q &\overset{\text{Keplerian}}{\equiv} \frac{\Omega c_s}{\pi G \Sigma}\\
    &\approx 2.83 \times 10^2 \pfrac{h/a}{5\%} \pfrac{m_{\star}}{m_{\odot}} \pfrac{\Sigma}{500\,\mathrm{g}\,\mathrm{cm}^{-2}}^{-1} \pfrac{a}{1\AU}^{-2}
  \end{aligned}
\end{equation}
is below unity. In an isothermal, hydrostatic and non-self-gravitating disk, the midplane density $\rho_0$ is linked to the surface density $\Sigma$ via $\rho_0 = \Sigma / h \sqrt{2\pi}$. We sample values of $Q\in \left[10^{1/2},10^2\right]$ by tuning the background density $\rho_0$. Whether realistic or not in the context of super-Earths formation \citep{chiang13}, such low values of $Q$ are required to reveal gas-gravity effects in the envelope. 

Given the gas density distribution, we solve Poisson's equation to obtain its gravitational potential. The specific method and its validation are presented in Appendix \ref{app:selfgravity}. Characteristics of self-gravitating simulations are listed in \autoref{tab:sgruns}, where the labels are appended by the corresponding value of $Q$. 

\begin{table}%[H]
\caption{List of isothermal self-gravitating simulations with their label, pressure scale $H$, Bondi radius $B$, Toomre parameter $Q$, and Bondi mass $m_{\mathrm{B}}$ relative to the core mass.}
\label{tab:sgruns}
\begin{tabular}{llcccr}
Label & $H$ & $B$ & $\log_{10}Q$ & $m_{\mathrm{B}} / m_c$\\
\hline
\verb!IH16B16Q100! & $16$ & $16$ & $2$ & $6.7\times 10^{-4}$ \\
\verb!IH16B16Q10! & $16$ & $16$ & $1$ & $7.2\times 10^{-3}$ \\
\verb!IH16B16Q3! & $16$ & $16$ & $1/2$ & $3.9\times 10^{-2}$ \\
\verb!IH32B4Q3! & $32$ & $4$ & $1/2$ & $3.2 \times 10^{-3}$\\
\verb!IH32B8Q100! & $32$ & $8$ & $2$ & $1.9 \times 10^{-4}$\\
\verb!IH32B8Q10! & $32$ & $8$ & $1$ & $2.1 \times 10^{-3}$ \\
\verb!IH32B8Q3! & $32$ & $8$ & $1/2$ & $7.0 \times 10^{-3}$ \\
\end{tabular}
\end{table}

%%%%%%%%%%%%%%%%%%%%%%%%%%%%%%%%%%%%%%%%%%%%%%%%%%%

\subsection{Self-gravitating envelopes}

In the limit $Q\rightarrow \infty$, the gravity of the gas is negligible compared to the gravity of the core, so the density distribution should scale linearly with the background density $\rho_0$. If we decrease $Q$ (increase $\rho_0$) while maintaining the distribution of $\rho/\rho_0$, then the mass of the envelope increases relative to the mass of the core and adds up to it. Self-gravitating envelopes are therefore expected to pull more gas toward the core and become more massive. 

We integrate the mass $m_{\mathrm{B}}$ inside the Bondi disk and find that it remains below $4\%$ of the mass of the core at the end of every two-dimensional simulation presented here (see \autoref{tab:sgruns}). To measure the contribution of the gas gravity, we compare $m_{\mathrm{B}}$ to the mass $m_0 \equiv \pi (\rB^2-r_c^2) \rho_0$ of the background disk over the same area. The ratio $m_{\mathrm{B}}/m_0$ increases by less than $30\%$ when decreasing $Q$ in our sample of simulations. Run \texttt{IH32B8Q3} has a density corresponding to $10^{-1/2}\approx 32\%$ of the critical density for gravitational fragmentation, which is $10^{3/2}\approx 32$ times larger than the density of run \texttt{IH32B8Q100}, yet the ratio of $m_{\mathrm{B}}/m_0$ is only $16\%$ larger. As in their non-self-gravitating analogues, runs \texttt{IH32B32Q100} and \texttt{IH32B32Q10} feature mass accretion onto the core due to steady shocks inside their rotating envelope. The mass accretion rate normalized by $\Omega m_0$ is $25\%$ larger in run \texttt{IH32B32Q10} compared to the less massive case \texttt{IH32B32Q100}. 

\begin{figure}%[H]
\begin{center}
\includegraphics[width=\columnwidth]{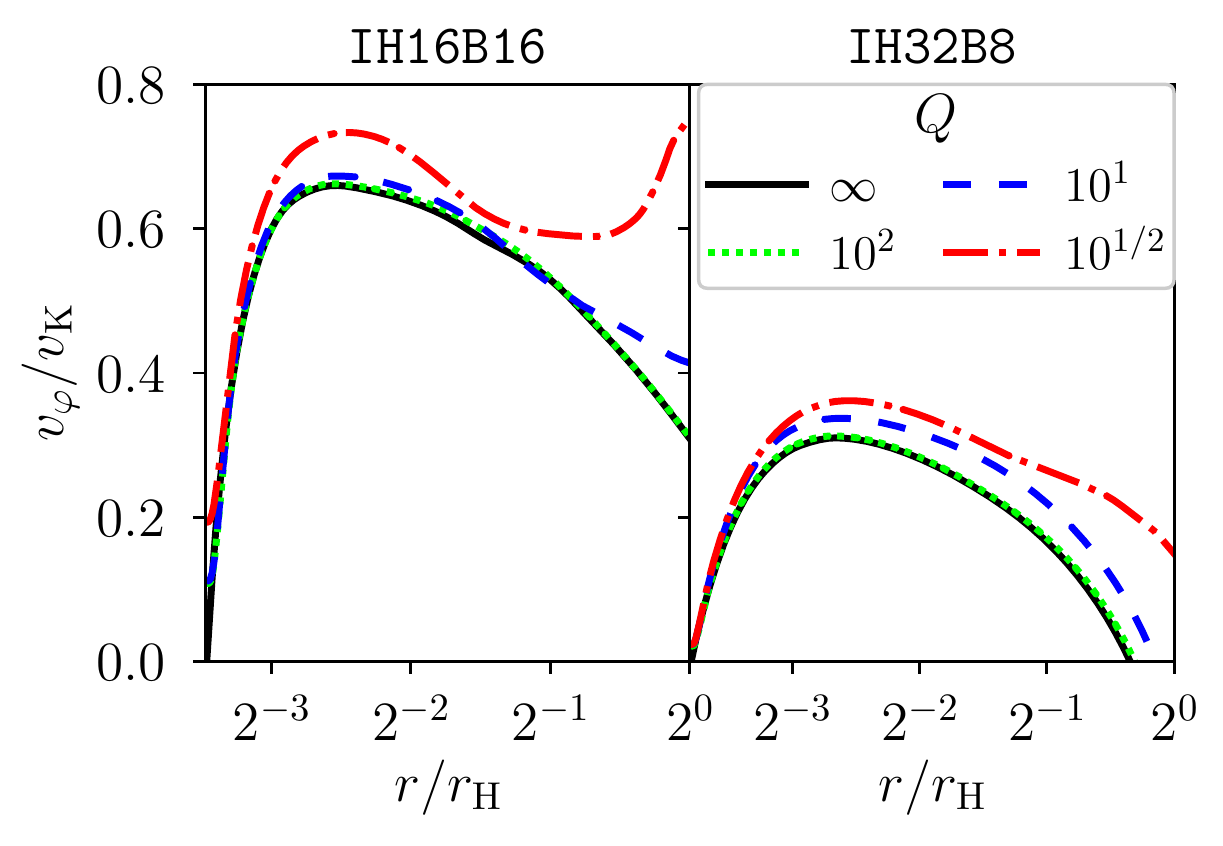}
\end{center}
\caption{Radial profiles of rotational support $v_{\varphi}/\vK$ in isothermal simulations with different values of the Toomre parameter $Q$ (see legend); \emph{left panel}: massive core with $H=B=16$; \emph{right panel}: low-mass core with $H=32$ and $B=8$. \label{fig:g2_vpp}}
\end{figure}

We previously demonstrated the importance of rotational support for the envelope dynamics. We now examine how the radial momentum balance of the envelope is affected by the gas self-gravity. We draw on \autoref{fig:g2_vpp} the radial profiles of $v_{\varphi}/\vK$ measured in the time-averaged flow of a series of self-gravitating simulations. Note that $\vK \equiv \sqrt{r\partial_r\Phi}$ includes the mass of the enclosed gas. We find that the degree of rotational support $v_{\varphi}/\vK$ decreases with $Q$, i.e. it increases with the mass of the envelope. The gas gravity affects the envelopes of low and high-mass cores similarly with respect to their rotational support. The ratio $v_{\varphi}/\vK$ increases by less than $10\%$ in every case considered. This is to be expected given the $<4\%$ envelope mass increase measured inside the Bondi radius. We conclude that for the range of core masses considered here, two-dimensional envelopes are only moderately affected by their self-gravity. 

%%%%%%%%%%%%%%%%%%%%%%%%%%%%%%%%%%%%%%%%%%%%%%%%%%%

\subsection{Self-gravitating density waves}

\begin{figure}%[H] 
\begin{center}
\includegraphics[width=\columnwidth]{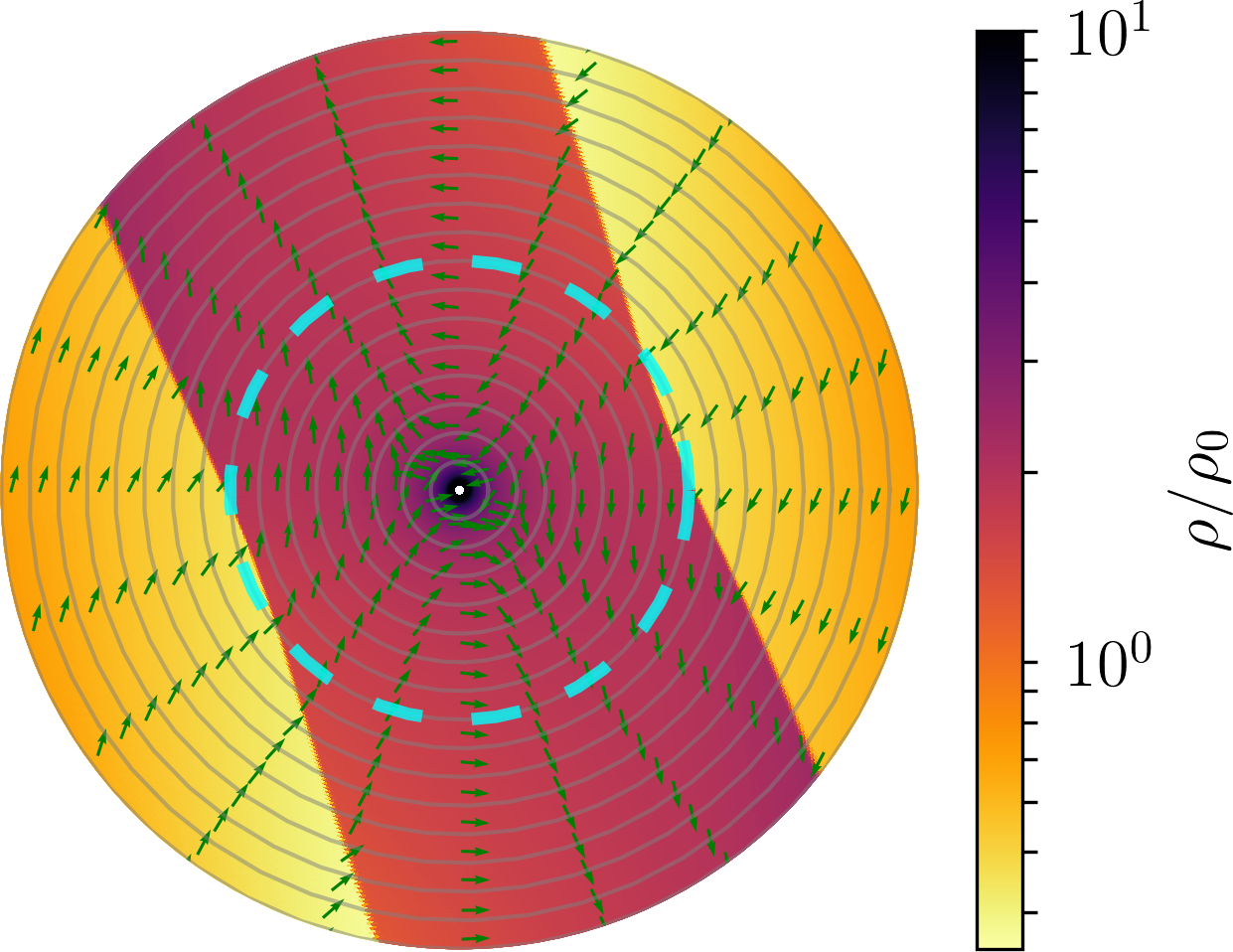}
\end{center}
\caption{Time-averaged flow in the self-gravitating simulation \texttt{IH32B4Q3}: the color map shows the density relative to the background density, and the arrows are tangent to the local velocity field; the circles mark every $4r_c$ (solid grey), the pressure scale $h$ (dashed cyan)and the Hill radius $\rH$ (dotted green). \label{fig:i2dp32b4q3_Map_tot}}
\end{figure}

We find that in our self-gravitating simulations the mass in the spiral density waves can become comparable to the mass enclosed inside two-dimensional envelopes. We illustrate this by showing the density distribution of run \texttt{IH32B4Q3} on \autoref{fig:i2dp32b4q3_Map_tot}. With $B=4$, the density at the surface of the core is only $26$ times larger than the background disk density, corresponding to $29\%$ of its hydrostatic value from \eqref{eqn:rhohsisot}. As previously, stationary shock waves are launched into the disk around one pressure scale $h$ away from the core. These shocks extend in the co-orbiting (horse-shoe) region beyond $2h$ upstream of the core. Such extended shocks appeared only around high-mass cores in non-self-gravitating simulations; see \autoref{fig:i2dp16b32_Map_wda} for example. The core of run \texttt{IH32B4Q3} is embedded in a homogeneously shocked medium, with circulating streamlines restricted to the innermost $r \lesssim \rH/2$. The rest of the shocked gas is recycled on orbital timescales. 

%The spiral waves cover a much larger area than the Bondi disk. The potential of the gas, solution of Poisson's equation $\Delta \Phi_g = 4\pi \rho$, is most sensitive to the large-scale density structures. We therefore expect the gas gravity to be dynamically important on larger scales $\gtrsim h$. 

\begin{figure}%[H] 
\begin{center}
\includegraphics[width=\columnwidth]{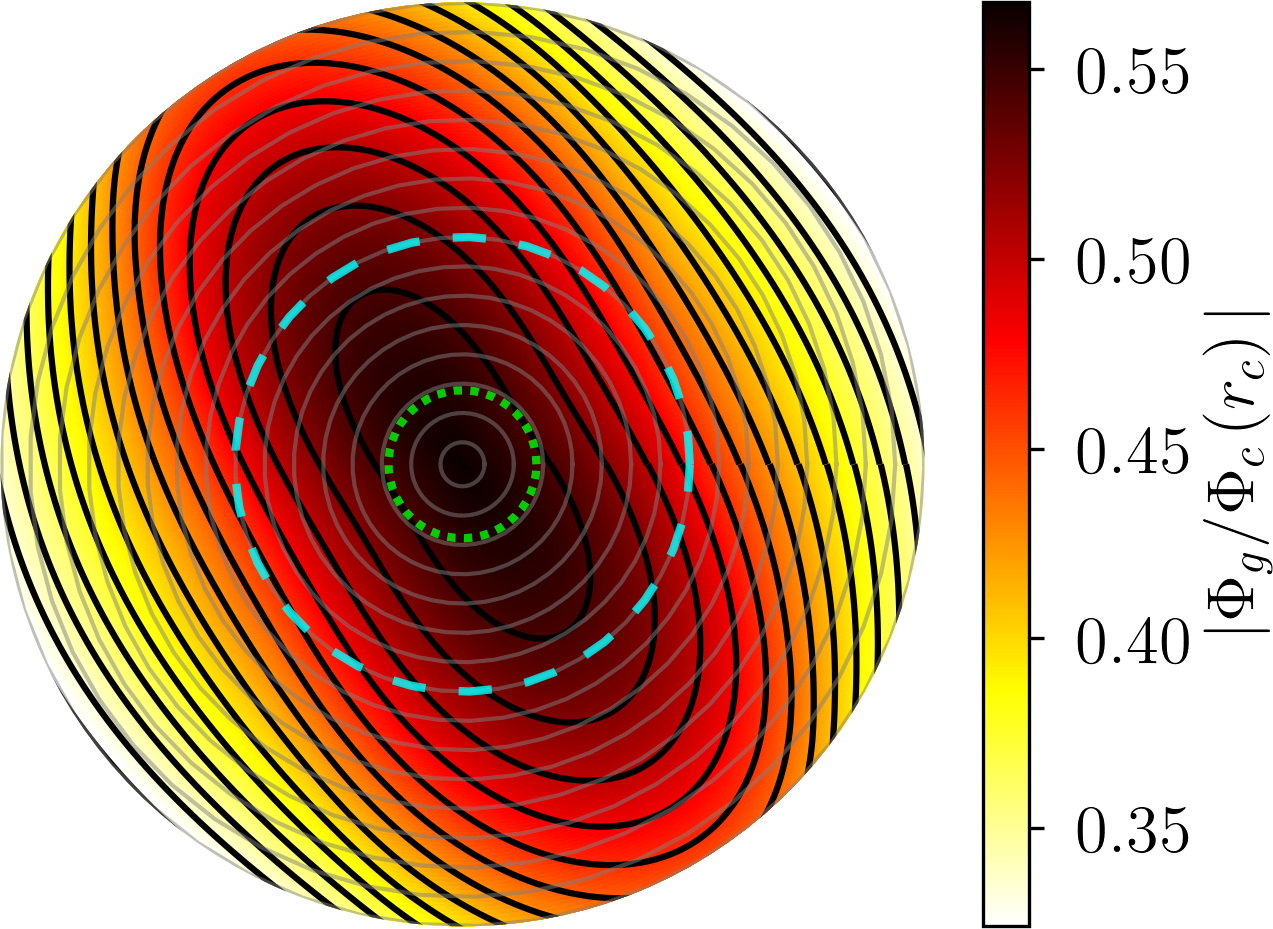}
\end{center}
\caption{Gravitational potential of the gas $\Phi_g$ relative to the potential of the core $\Phi_c(r_c)$ in run \texttt{IH32B4Q3}, corresponding to the density distribution shown on \autoref{fig:i2dp32b4q3_Map_tot}. Both potentials reach approximately zero at the outer radial boundary. \label{fig:i2dp32b4q3_Map_phig}} 
\end{figure}

Given the density distribution of \autoref{fig:i2dp32b4q3_Map_tot}, the gravitational potential $\Phi_g$ solution of Poisson's equation is represented on \autoref{fig:i2dp32b4q3_Map_phig}. The iso-potential contours are elongated along the axis of the shocked and dense region. The potential of the gas remains deeper than $50\%$ of $\Phi_c(r_c)$ over an area several times larger than the Hill disk. The potential of the gas effectively dominates over the potential of the core outside the Hill radius, competing with the tidal potential of the star. Run \texttt{IH32B4Q3} thus illustrates how a low-mass core can launch self-gravitating density waves into a massive protoplanetary disk. The density perturbations, although not as pronounced as around massive cores, are amplified by their self-gravity beyond one pressure scale away from the core. With stronger shocks spanning a larger area, self-gravity could significantly alter the observational signatures of even the relatively low-mass planetary cores \citep{dong15}.

%%%%%%%%%%%%%%%%%%%%%%%%%%%%%%%%%%%%%%%%%%%%%%%%%%%
%%%%%%%%%%%%%%%%%%%%%%%%%%%%%%%%%%%%%%%%%%%%%%%%%%%

\section{Discussion} 
\label{sec:discussion}

%%%%%%%%%%%%%%%%%%%%%%%%%%%%%%%%%%%%%%%%%%%%%%%%%%%

%%%%%%%%%%%%%%%%%%%%%%%%%%%%%%%%%%%%%%%%%%%%%%%%%%%

\subsection{Comparison with previous studies}

%%%%%%%%%%%%%%%%%%%%%%%%%%%%%%%%%%%%%%%%%%%%%%%%%%%

Despite the difference in computational power, our simulations reproduce most of the flow properties already identified by \cite{miki82} in two-dimensional inviscid simulations. The flow pattern can be divided into three distinct regions: the background shear, the co-orbital (horse-shoe) flow, and an inner region where streamlines circulate around the core. As a result of vortensity conservation, the innermost regions become rotationally supported (with a prograde orientation) when the envelope mass increases. 
%The inertia of the rotating envelope can be comparable to its pressure support against the gravity of the core. 
The radial momentum balance of envelopes forming around massive cores may therefore be far from hydrostatic, contrary to the usual assumption of one-dimensional models \citep[e.g.,][]{mizuno78,pollack96,rafikov06}. 

Contrarily to \cite{kley99} and \cite{npmk00}, we do not include any explicit viscosity and we do not remove mass at the location of the core. Our setup allows steady envelopes to form with no mass accretion toward the core. However, we do observe mass accretion in a subset of simulations with massive cores, satisfying $\rH/h\gtrsim 1$ ($m_c\gtrsim m_{\rm th}$). Shocks form in this regime \citep{KP96,lubow99} on scales previously under-resolved. Shocks break vortensity conservation. The sign of the vortensity jump depends on the geometry of the shock \citep{kevlahan97}; we find that the inner envelope generally loses vortensity. With a reduced vortensity, the envelope is less rotationally supported. To maintain radial momentum balance, the envelope must increase its pressure support by accreting mass, so shocks effectively lower the `centrifugal barrier' around sufficiently massive cores in two dimensions \citep{ormel1}. In the low-mass regime $B/H<1$, our results agree with those of \cite{ormel1} in terms of vortensity conservation and resulting rotation profiles (see \autoref{sec:circarcore} below).

In addition to momentum dissipation, shocks drive temporal variability in the envelope. The `turbulent' mass flux is always small compared to the laminar one, but it allows the mixing of material from the different regions of the flow. 

Recycling the envelope material with fresh gas from the disk is a possible way to counteract the radiative cooling of the envelope \citep{ormel2}. We postpone the discussion of recycling to later three-dimensional results, noting that the strength of the shocks should be significantly reduced in three-dimensions \citep[e.g., ][]{bate03}. 

The main caveat of our 2D model is the local approximation. Planets in the parameter range $B/H \gtrsim 1$ are expected to open a gap in the disk \citep[][]{kley12}, which we cannot capture without ad-hoc boundary conditions. Global models are also necessary to capture the accumulation of vortensity in the co-orbiting flow, which could affect the migration rate of the planet \citep{ward91,koller03,PP09} and that of dust grains through the orbit of the planet \citep{weidenschilling77}. 

%%%%%%%%%%%%%%%%%%%%%%%%%%%%%%%%%%%%%%%%%%%%%%%%%%%

\subsection{Global sensitivity to the core radius} 
\label{sec:circarcore}

%%%%%%%%%%%%%%%%%%%%%%%%%%%%%%%%%%%%%%%%%%%%%%%%%%%

Results of \autoref{sec:isotsimu} demonstrate the sensitivity of the flow structure to the core size $r_c$. To gain insight into the origin of this dependence, we analyze a simplified model for the envelope structure to examine its dependence on the size of the core. As mentioned in \autoref{sec:vortensity}, in barotropic flows the vortensity $\varpi_z \equiv \left(\omega_z+2\Omega\right)/\rho$ is constant along streamlines. Assuming that the $\varpi_z$ has kept its initial value $\varpi_0$ everywhere, we integrate the vorticity flux over the disk of radius $r$ around the core and apply Stokes theorem:
\begin{subequations} 
\label{eqn:vortens}
\begin{align} 
  \Gamma &\equiv \int_{\mathcal{D}(r)}\!\!\!\!\!\! \left( \nabla\times v + 2\Omega \right) \cdot \dd S \simeq \pi r^2 \varpi_0 \brac{\rho}_{\mathcal{D}} \label{eqn:vortens1}\\
  &=\oint_{\mathcal{C}(r)}\!\!\!\!\!\! v\cdot \dd \ell + 2\Omega \pi (r^2-r_c^2) \simeq 2\pi r \brac{v_{\varphi} + \Omega r}_{\mathcal{C}}. \label{eqn:vortens2}
\end{align}
\end{subequations}
We have neglected the contribution from the inner radial boundary at $r=r_c$ and assumed a purely circular flow in \eqref{eqn:vortens2}. This relation links the velocity on a contour to the mass enclosed inside this contour. It is accurately satisfied in our simulations, from the inner radius up to the envelope boundary at $r \simeq h$. In the hydrostatic limit, the mass contained in the isothermal envelope $\pi r^2 \brac{\rho}_{\mathcal{D}}$ diverges when the inner radius $r_c \rightarrow 0$. The amount of rotational support in such an envelope should therefore depend on the size of the core. 

The azimuthal velocity $v_{\varphi}$ must increase with the mass of the envelope, eventually becoming positive \citep[prograde,][]{miki82}. Assuming axisymmetry, we can compute the structure of an isothermal envelope in radial momentum balance: 
\begin{align}
  \frac{\partial \log \rho}{\partial \log r} &= \frac{v_{\varphi}^2}{c_s^2} +  \frac{2\Omega r}{c_s} \frac{v_{\varphi}}{c_s} - \frac{\rB}{r}, \label{eqn:radmessint1}\\
  \frac{1}{r}\partial_r\left[r v_{\varphi}\right] &= \varpi_z \rho - 2\Omega, \label{eqn:radmessint2}
\end{align}
where $\varpi_z$ is the $z$-component of the vortensity, assumed to be fixed. \cite{ormel1} examined the same system omitting the Coriolis acceleration in \eqref{eqn:radmessint1}. The Coriolis term is always negligible around rotationally supported envelopes, but we keep it for the sake of completeness. 

We solve this system numerically by imposing $v_{\varphi}(r_c)=0$, $\rho(h)=1$, and the Keplerian vortensity $\varpi_z=\varpi_0\equiv\Omega/2\rho_0$ everywhere. The solutions are in good agreement with those obtained from direct simulations (see \autoref{fig:i2dp16b16_SH_staticrhovp}). Our purpose here is not to reproduce exactly the two-dimensional results, but to capture the main features of the equilibrium, which do not depend on the exact choice of $\rho(h)$. Given the density and velocity profiles, we can compare the pressure to the inertial support against gravity at every radius in \eqref{eqn:radmombal}. We focus on the profiles of $v_{\varphi}/\vK$ in the prograde ($v_{\varphi}>0$) portion of the envelope; the angular velocity $v_{\varphi}$ smoothly takes negative values outside of this inner region.

\begin{figure}%[H]
\begin{center}
\includegraphics[width=1.0\columnwidth]{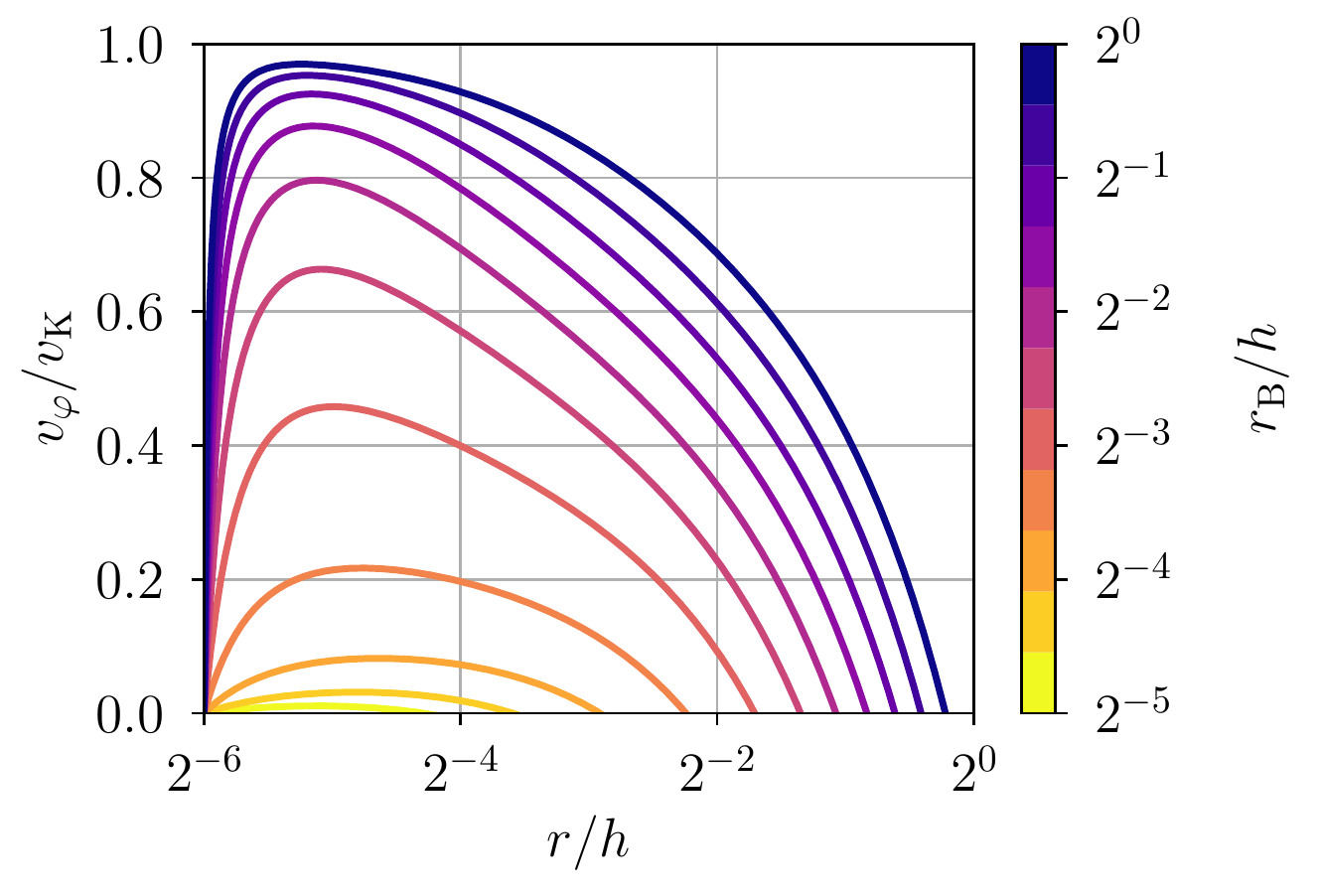}
\caption{Rotational support $v_{\varphi}/\vK$ solution of \eqref{eqn:radmessint1}-\eqref{eqn:radmessint2} for $H=64$, each curve corresponding to a different $\rB/h\leq 1$. \label{fig:mappi_psh64}}
\end{center}
\end{figure}

\begin{figure}%[H]
\begin{center}
\includegraphics[width=1.0\columnwidth]{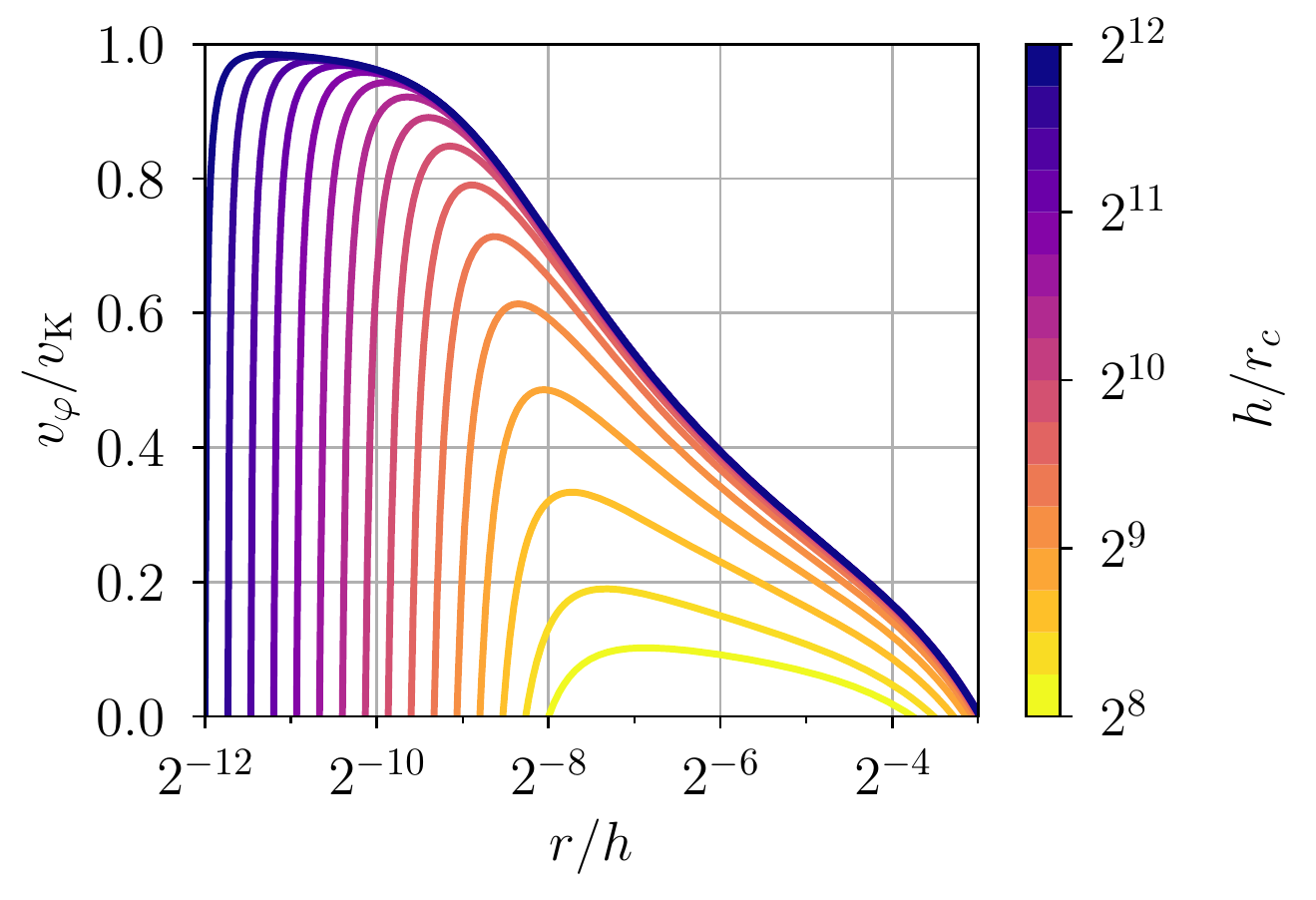}
\caption{Rotational support $v_{\varphi}/\vK$ solution of \eqref{eqn:radmessint1}-\eqref{eqn:radmessint2} for $B/H=2.4\times 10^{-2}$ and different values of $H\equiv h/r_c$. \label{fig:super_boh0024}}
\end{center}
\end{figure}

First, we set the core size via $H\equiv h/r_c = 64$ according to \eqref{eqn:magnH} and we compute solutions for different core masses by varying $\rB/h$. This family of solutions is represented on \autoref{fig:mappi_psh64}. The amount of rotational support, as measured by $v_{\varphi}/\vK$, increases in the entire envelope as a function of $\rB/h$. In this case, the threshold $v_{\varphi}/\vK=50\%$ is attained for $B\equiv \rB/r_c \gtrsim 8$. The radial extent of the prograde envelope also increases with $\rB/h$ and approaches the outer boundary $r=h$ as $\rB/h$ increases. Solutions with $\rB/h>1$ are not represented because they become excessively stiff and require higher resolutions. 

Next, we set the ratio of $B/H = 2.4\times 10^{-2}$ as estimated in \eqref{eqn:magnBoH}, and we vary the core radius relative to the local pressure scale via $H$. With this parametrization, the ratio $\rH/h$ remains constant, the mass of the core increases as $m_c \propto H^3$ so the Keplerian velocity $\vK \propto H^{3/2}$. This family of solutions is represented on \autoref{fig:super_boh0024}. The envelope becomes more rotationally supported as $H$ increases, i.e. as the core radius $r_c$ decreases with respect to $h$. All the simulations presented in this paper fall in this regime sensitive to the core radius. However, the curves appear to converge to an asymptotic profile of $v_{\varphi}/\vK$ over the whole envelope. We conclude that for sufficiently small cores, the amount of rotational support in the envelope should eventually become independent of the core radius.

%%%%%%%%%%%%%%%%%%%%%%%%%%%%%%%%%%%%%%%%%%%%%%%%%%%
%%%%%%%%%%%%%%%%%%%%%%%%%%%%%%%%%%%%%%%%%%%%%%%%%%%

\section{Summary} 
\label{sec:summary}

%%%%%%%%%%%%%%%%%%%%%%%%%%%%%%%%%%%%%%%%%%%%%%%%%%%

We performed two-dimensional inviscid hydrodynamic simulations of embedded protoplanetary cores in a local model of Keplerian disk. We focused on the properties of the dense envelope surrounding the core in a regime where its Bondi radius spans across the hydrostatic pressure scale of the disk (i.e. as the core mass is varied from sub-thermal to super-thermal). The core was included as a spatially resolved boundary, and gas was treated as either isothermal or adiabatic. We also implemented a Poisson solver in the \textsc{pluto} code in order to include the gas self-gravity in a subset of simulations. Our main conclusions are the following. 
\begin{enumerate}

\item Vortensity is conserved in two-dimensional flows around low (sub-thermal) mass cores, so the amount of rotational support in the envelope depends on the mass inside the envelope and, a priori, on the size of the core. This dependence is expected to vanish in the limit of cores small relative to the pressure scale of the disk, which is computationally more challenging to simulate. 

\item Stationary shocks form inside the envelope of massive cores, when the circulating flow becomes supersonic. By altering the vortensity distribution, these shocks allow mass accretion and effectively break the centrifugal barrier. They also drive a turbulent mixing of the envelope material with the background disk. Stationary shocks also form in the co-orbital flow, far upstream of massive cores. 

\item Adiabatic envelopes are more pressure-supported than their isothermal analogues; they are more efficiently recycled around low-mass cores but less susceptible to shock-induced mixing in the high-mass regime. The shock-induced (irreversible) heating is small compared to the adiabatic heating due to gas compression in every case studied. 

\item The self-gravitating potential of the gas is most sensitive to the large-scale density waves launched into the disk. Self-gravity affects primarily the extent and intensity of these large-scale shock waves, whereas the properties of the inner envelope only weakly depend on the Toomre parameter $Q$ of the disk. 
\end{enumerate}

The extension of this study to three-dimensions, with a focus on the recycling properties of the flow, will be the subject of the subsequent work.

\section*{Acknowledgements}

We thank the anonymous referee for providing constructive comments that improved the clarity of this paper. Financial support of this work by Isaac Newton Trust, Department of Applied Mathematics and Theoretical Physics and STFC through grant ST/P000673/1 is gratefully acknowledged.

%%%%%%%%%%%%%%%%%%%%%%%%%%%%%%%%%%%%%%%%%%%%%%%%%%



\bibliographystyle{mnras}
\bibliography{biblio} % if your bibtex file is called example.bib

\begin{thebibliography}{}
\makeatletter
\relax
\def\mn@urlcharsother{\let\do\@makeother \do\$\do\&\do\#\do\^\do\_\do\%\do\~}
\def\mn@doi{\begingroup\mn@urlcharsother \@ifnextchar [ {\mn@doi@}
  {\mn@doi@[]}}
\def\mn@doi@[#1]#2{\def\@tempa{#1}\ifx\@tempa\@empty \href
  {http://dx.doi.org/#2} {doi:#2}\else \href {http://dx.doi.org/#2} {#1}\fi
  \endgroup}
\def\mn@eprint#1#2{\mn@eprint@#1:#2::\@nil}
\def\mn@eprint@arXiv#1{\href {http://arxiv.org/abs/#1} {{\tt arXiv:#1}}}
\def\mn@eprint@dblp#1{\href {http://dblp.uni-trier.de/rec/bibtex/#1.xml}
  {dblp:#1}}
\def\mn@eprint@#1:#2:#3:#4\@nil{\def\@tempa {#1}\def\@tempb {#2}\def\@tempc
  {#3}\ifx \@tempc \@empty \let \@tempc \@tempb \let \@tempb \@tempa \fi \ifx
  \@tempb \@empty \def\@tempb {arXiv}\fi \@ifundefined
  {mn@eprint@\@tempb}{\@tempb:\@tempc}{\expandafter \expandafter \csname
  mn@eprint@\@tempb\endcsname \expandafter{\@tempc}}}

\bibitem[\protect\citeauthoryear{Ayliffe \& Bate}{Ayliffe \&
  Bate}{2009}]{ayliffe09}
Ayliffe B.~A.,  Bate M.~R.,  2009, \mn@doi [Monthly Notices of the Royal
  Astronomical Society] {10.1111/j.1365-2966.2009.15002.x}, 397, 657

\bibitem[\protect\citeauthoryear{Balay, Gropp, McInnes  \& Smith}{Balay
  et~al.}{1997}]{petsc-efficient}
Balay S.,  Gropp W.~D.,  McInnes L.~C.,   Smith B.~F.,  1997, in Arge E.,
  Bruaset A.~M.,   Langtangen H.~P.,  eds, Modern Software Tools in Scientific
  Computing. Birkh{\"{a}}user Press, pp 163--202

\bibitem[\protect\citeauthoryear{Balay et~al.,}{Balay
  et~al.}{2018}]{petsc-user-ref}
Balay S.,  et~al., 2018, Technical Report ANL-95/11 - Revision 3.9, {PETS}c
  Users Manual, \url {http://www.mcs.anl.gov/petsc}.
Argonne National Laboratory, \url {http://www.mcs.anl.gov/petsc}

\bibitem[\protect\citeauthoryear{{Batalha} et~al.,}{{Batalha}
  et~al.}{2013}]{batalha13}
{Batalha} N.~M.,  et~al., 2013, \mn@doi [\apjs] {10.1088/0067-0049/204/2/24},
  \href {http://adsabs.harvard.edu/abs/2013ApJS..204...24B} {204, 24}

\bibitem[\protect\citeauthoryear{{Bate}, {Lubow}, {Ogilvie}  \&
  {Miller}}{{Bate} et~al.}{2003}]{bate03}
{Bate} M.~R.,  {Lubow} S.~H.,  {Ogilvie} G.~I.,   {Miller} K.~A.,  2003,
  \mn@doi [\mnras] {10.1046/j.1365-8711.2003.06406.x}, \href
  {http://adsabs.harvard.edu/abs/2003MNRAS.341..213B} {341, 213}

\bibitem[\protect\citeauthoryear{{Chachan} \& {Stevenson}}{{Chachan} \&
  {Stevenson}}{2018}]{chachanstevenson18}
{Chachan} Y.,  {Stevenson} D.~J.,  2018, \mn@doi [\apj]
  {10.3847/1538-4357/aaa459}, \href
  {http://adsabs.harvard.edu/abs/2018ApJ...854...21C} {854, 21}

\bibitem[\protect\citeauthoryear{{Chiang} \& {Laughlin}}{{Chiang} \&
  {Laughlin}}{2013}]{chiang13}
{Chiang} E.,  {Laughlin} G.,  2013, \mn@doi [\mnras] {10.1093/mnras/stt424},
  \href {http://adsabs.harvard.edu/abs/2013MNRAS.431.3444C} {431, 3444}

\bibitem[\protect\citeauthoryear{D'Angelo \& Bodenheimer}{D'Angelo \&
  Bodenheimer}{2013}]{dangelo13}
D'Angelo G.,  Bodenheimer P.,  2013, \mn@doi [The Astrophysical Journal]
  {10.1088/0004-637X/778/1/77}, 778, 77

\bibitem[\protect\citeauthoryear{{Dong}, {Rafikov}  \& {Stone}}{{Dong}
  et~al.}{2011}]{dong11}
{Dong} R.,  {Rafikov} R.~R.,   {Stone} J.~M.,  2011, \mn@doi [\apj]
  {10.1088/0004-637X/741/1/57}, \href
  {http://adsabs.harvard.edu/abs/2011ApJ...741...57D} {741, 57}

\bibitem[\protect\citeauthoryear{{Dong}, {Zhu}, {Rafikov}  \& {Stone}}{{Dong}
  et~al.}{2015}]{dong15}
{Dong} R.,  {Zhu} Z.,  {Rafikov} R.~R.,   {Stone} J.~M.,  2015, \mn@doi [\apjl]
  {10.1088/2041-8205/809/1/L5}, \href
  {http://adsabs.harvard.edu/abs/2015ApJ...809L...5D} {809, L5}

\bibitem[\protect\citeauthoryear{{Fressin} et~al.,}{{Fressin}
  et~al.}{2013}]{fressin13}
{Fressin} F.,  et~al., 2013, \mn@doi [\apj] {10.1088/0004-637X/766/2/81}, \href
  {http://adsabs.harvard.edu/abs/2013ApJ...766...81F} {766, 81}

\bibitem[\protect\citeauthoryear{{Fung} \& {Lee}}{{Fung} \&
  {Lee}}{2018}]{funglee18}
{Fung} J.,  {Lee} E.~J.,  2018, \mn@doi [\apj] {10.3847/1538-4357/aabaf7},
  \href {http://adsabs.harvard.edu/abs/2018ApJ...859..126F} {859, 126}

\bibitem[\protect\citeauthoryear{{Ginzburg} \& {Sari}}{{Ginzburg} \&
  {Sari}}{2017}]{ginzburg17}
{Ginzburg} S.,  {Sari} R.,  2017, \mn@doi [\mnras] {10.1093/mnras/stw2637},
  \href {http://adsabs.harvard.edu/abs/2017MNRAS.464.3937G} {464, 3937}

\bibitem[\protect\citeauthoryear{{Ginzburg}, {Schlichting}  \&
  {Sari}}{{Ginzburg} et~al.}{2016}]{ginzburg16}
{Ginzburg} S.,  {Schlichting} H.~E.,   {Sari} R.,  2016, \mn@doi [\apj]
  {10.3847/0004-637X/825/1/29}, \href
  {http://adsabs.harvard.edu/abs/2016ApJ...825...29G} {825, 29}

\bibitem[\protect\citeauthoryear{{Ginzburg}, {Schlichting}  \&
  {Sari}}{{Ginzburg} et~al.}{2018}]{ginzburg18}
{Ginzburg} S.,  {Schlichting} H.~E.,   {Sari} R.,  2018, \mn@doi [\mnras]
  {10.1093/mnras/sty290}, \href
  {http://adsabs.harvard.edu/abs/2018MNRAS.476..759G} {476, 759}

\bibitem[\protect\citeauthoryear{{Goodman} \& {Rafikov}}{{Goodman} \&
  {Rafikov}}{2001}]{goodfikov01}
{Goodman} J.,  {Rafikov} R.~R.,  2001, \mn@doi [\apj] {10.1086/320572}, \href
  {http://adsabs.harvard.edu/abs/2001ApJ...552..793G} {552, 793}

\bibitem[\protect\citeauthoryear{{Gorti}, {Liseau}, {S{\'a}ndor}  \&
  {Clarke}}{{Gorti} et~al.}{2016}]{gorti16}
{Gorti} U.,  {Liseau} R.,  {S{\'a}ndor} Z.,   {Clarke} C.,  2016, \mn@doi
  [\ssr] {10.1007/s11214-015-0228-x}, \href
  {http://adsabs.harvard.edu/abs/2016SSRv..205..125G} {205, 125}

\bibitem[\protect\citeauthoryear{Hill}{Hill}{1878}]{hill78}
Hill G.~W.,  1878, \mn@doi [American journal of Mathematics] {10.2307/2369430},
  1, 5

\bibitem[\protect\citeauthoryear{{Howard} et~al.,}{{Howard}
  et~al.}{2012}]{howard12}
{Howard} A.~W.,  et~al., 2012, \mn@doi [\apjs] {10.1088/0067-0049/201/2/15},
  \href {http://adsabs.harvard.edu/abs/2012ApJS..201...15H} {201, 15}

\bibitem[\protect\citeauthoryear{{Inamdar} \& {Schlichting}}{{Inamdar} \&
  {Schlichting}}{2016}]{Inamdar15}
{Inamdar} N.~K.,  {Schlichting} H.~E.,  2016, \mn@doi [\apjl]
  {10.3847/2041-8205/817/2/L13}, \href
  {http://adsabs.harvard.edu/abs/2016ApJ...817L..13I} {817, L13}

\bibitem[\protect\citeauthoryear{Kevlahan}{Kevlahan}{1997}]{kevlahan97}
Kevlahan N.-R.,  1997, Journal of Fluid Mechanics, 341, 371

\bibitem[\protect\citeauthoryear{{Kley}}{{Kley}}{1998}]{kley98}
{Kley} W.,  1998, \aap, \href
  {http://adsabs.harvard.edu/abs/1998A%26A...338L..37K} {338, L37}

\bibitem[\protect\citeauthoryear{{Kley}}{{Kley}}{1999}]{kley99}
{Kley} W.,  1999, \mn@doi [\mnras] {10.1046/j.1365-8711.1999.02198.x}, \href
  {http://adsabs.harvard.edu/abs/1999MNRAS.303..696K} {303, 696}

\bibitem[\protect\citeauthoryear{{Kley} \& {Nelson}}{{Kley} \&
  {Nelson}}{2012}]{kley12}
{Kley} W.,  {Nelson} R.~P.,  2012, \mn@doi [\araa]
  {10.1146/annurev-astro-081811-125523}, \href
  {http://adsabs.harvard.edu/abs/2012ARA%26A..50..211K} {50, 211}

\bibitem[\protect\citeauthoryear{{Koller}, {Li}  \& {Lin}}{{Koller}
  et~al.}{2003}]{koller03}
{Koller} J.,  {Li} H.,   {Lin} D.~N.~C.,  2003, \mn@doi [\apjl]
  {10.1086/379032}, \href {http://adsabs.harvard.edu/abs/2003ApJ...596L..91K}
  {596, L91}

\bibitem[\protect\citeauthoryear{{Korycansky} \& {Papaloizou}}{{Korycansky} \&
  {Papaloizou}}{1996}]{KP96}
{Korycansky} D.~G.,  {Papaloizou} J.~C.~B.,  1996, \mn@doi [\apjs]
  {10.1086/192311}, \href {http://adsabs.harvard.edu/abs/1996ApJS..105..181K}
  {105, 181}

\bibitem[\protect\citeauthoryear{{Kusaka}, {Nakano}  \& {Hayashi}}{{Kusaka}
  et~al.}{1970}]{kusaka70}
{Kusaka} T.,  {Nakano} T.,   {Hayashi} C.,  1970, \mn@doi [Progress of
  Theoretical Physics] {10.1143/PTP.44.1580}, \href
  {http://adsabs.harvard.edu/abs/1970PThPh..44.1580K} {44, 1580}

\bibitem[\protect\citeauthoryear{{Lambrechts} \& {Johansen}}{{Lambrechts} \&
  {Johansen}}{2012}]{lambrechts12}
{Lambrechts} M.,  {Johansen} A.,  2012, \mn@doi [\aap]
  {10.1051/0004-6361/201219127}, \href
  {http://adsabs.harvard.edu/abs/2012A%26A...544A..32L} {544, A32}

\bibitem[\protect\citeauthoryear{{Lee} \& {Chiang}}{{Lee} \&
  {Chiang}}{2015}]{leechiang15}
{Lee} E.~J.,  {Chiang} E.,  2015, \mn@doi [\apj] {10.1088/0004-637X/811/1/41},
  \href {http://adsabs.harvard.edu/abs/2015ApJ...811...41L} {811, 41}

\bibitem[\protect\citeauthoryear{{Lee}, {Chiang}  \& {Ormel}}{{Lee}
  et~al.}{2014}]{leechiang14}
{Lee} E.~J.,  {Chiang} E.,   {Ormel} C.~W.,  2014, \mn@doi [\apj]
  {10.1088/0004-637X/797/2/95}, \href
  {http://adsabs.harvard.edu/abs/2014ApJ...797...95L} {797, 95}

\bibitem[\protect\citeauthoryear{{Lopez} \& {Fortney}}{{Lopez} \&
  {Fortney}}{2014}]{lopezfortney14}
{Lopez} E.~D.,  {Fortney} J.~J.,  2014, \mn@doi [\apj]
  {10.1088/0004-637X/792/1/1}, \href
  {http://adsabs.harvard.edu/abs/2014ApJ...792....1L} {792, 1}

\bibitem[\protect\citeauthoryear{{Lubow}, {Seibert}  \& {Artymowicz}}{{Lubow}
  et~al.}{1999}]{lubow99}
{Lubow} S.~H.,  {Seibert} M.,   {Artymowicz} P.,  1999, \mn@doi [\apj]
  {10.1086/308045}, \href {http://adsabs.harvard.edu/abs/1999ApJ...526.1001L}
  {526, 1001}

\bibitem[\protect\citeauthoryear{Machida, Kokubo, Inutsuka  \&
  Matsumoto}{Machida et~al.}{2010}]{machida10}
Machida M.~N.,  Kokubo E.,  Inutsuka S.-i.,   Matsumoto T.,  2010, \mn@doi
  [Monthly Notices of the Royal Astronomical Society]
  {10.1111/j.1365-2966.2010.16527.x}, 405, 1227

\bibitem[\protect\citeauthoryear{{Matsuo}, {Shibai}, {Ootsubo}  \&
  {Tamura}}{{Matsuo} et~al.}{2007}]{matsuo07}
{Matsuo} T.,  {Shibai} H.,  {Ootsubo} T.,   {Tamura} M.,  2007, \mn@doi [\apj]
  {10.1086/517964}, \href {http://adsabs.harvard.edu/abs/2007ApJ...662.1282M}
  {662, 1282}

\bibitem[\protect\citeauthoryear{{Mignone}, {Bodo}, {Massaglia}, {Matsakos},
  {Tesileanu}, {Zanni}  \& {Ferrari}}{{Mignone} et~al.}{2007}]{mignone07}
{Mignone} A.,  {Bodo} G.,  {Massaglia} S.,  {Matsakos} T.,  {Tesileanu} O.,
  {Zanni} C.,   {Ferrari} A.,  2007, \mn@doi [\apjs] {10.1086/513316}, \href
  {http://adsabs.harvard.edu/abs/2007ApJS..170..228M} {170, 228}

\bibitem[\protect\citeauthoryear{{Mignone}, {Flock}, {Stute}, {Kolb}  \&
  {Muscianisi}}{{Mignone} et~al.}{2012}]{mignone12}
{Mignone} A.,  {Flock} M.,  {Stute} M.,  {Kolb} S.~M.,   {Muscianisi} G.,
  2012, \mn@doi [\aap] {10.1051/0004-6361/201219557}, \href
  {http://adsabs.harvard.edu/abs/2012A%26A...545A.152M} {545, A152}

\bibitem[\protect\citeauthoryear{{Miki}}{{Miki}}{1982}]{miki82}
{Miki} S.,  1982, \mn@doi [Progress of Theoretical Physics]
  {10.1143/PTP.67.1053}, \href
  {http://adsabs.harvard.edu/abs/1982PThPh..67.1053M} {67, 1053}

\bibitem[\protect\citeauthoryear{{Mizuno}}{{Mizuno}}{1980}]{mizuno80}
{Mizuno} H.,  1980, \mn@doi [Progress of Theoretical Physics]
  {10.1143/PTP.64.544}, \href
  {http://adsabs.harvard.edu/abs/1980PThPh..64..544M} {64, 544}

\bibitem[\protect\citeauthoryear{{Mizuno}, {Nakazawa}  \& {Hayashi}}{{Mizuno}
  et~al.}{1978}]{mizuno78}
{Mizuno} H.,  {Nakazawa} K.,   {Hayashi} C.,  1978, \mn@doi [Progress of
  Theoretical Physics] {10.1143/PTP.60.699}, \href
  {http://adsabs.harvard.edu/abs/1978PThPh..60..699M} {60, 699}

\bibitem[\protect\citeauthoryear{{Mordasini}, {Alibert}, {Georgy}, {Dittkrist},
  {Klahr}  \& {Henning}}{{Mordasini} et~al.}{2012}]{mordasini12}
{Mordasini} C.,  {Alibert} Y.,  {Georgy} C.,  {Dittkrist} K.-M.,  {Klahr} H.,
  {Henning} T.,  2012, \mn@doi [\aap] {10.1051/0004-6361/201118464}, \href
  {http://adsabs.harvard.edu/abs/2012A%26A...547A.112M} {547, A112}

\bibitem[\protect\citeauthoryear{{M{\"u}ller}, {Kley}  \& {Meru}}{{M{\"u}ller}
  et~al.}{2012}]{muellerkleymeru12}
{M{\"u}ller} T.~W.~A.,  {Kley} W.,   {Meru} F.,  2012, \mn@doi [\aap]
  {10.1051/0004-6361/201118737}, \href
  {http://adsabs.harvard.edu/abs/2012A%26A...541A.123M} {541, A123}

\bibitem[\protect\citeauthoryear{{Nakagawa}, {Hayashi}  \&
  {Nakazawa}}{{Nakagawa} et~al.}{1983}]{nakagawa83}
{Nakagawa} Y.,  {Hayashi} C.,   {Nakazawa} K.,  1983, \mn@doi [\icarus]
  {10.1016/0019-1035(83)90234-8}, \href
  {http://adsabs.harvard.edu/abs/1983Icar...54..361N} {54, 361}

\bibitem[\protect\citeauthoryear{{Nelson}, {Papaloizou}, {Masset}  \&
  {Kley}}{{Nelson} et~al.}{2000}]{npmk00}
{Nelson} R.~P.,  {Papaloizou} J.~C.~B.,  {Masset} F.,   {Kley} W.,  2000,
  \mn@doi [\mnras] {10.1046/j.1365-8711.2000.03605.x}, \href
  {http://adsabs.harvard.edu/abs/2000MNRAS.318...18N} {318, 18}

\bibitem[\protect\citeauthoryear{{Ormel} \& {Klahr}}{{Ormel} \&
  {Klahr}}{2010}]{ormelklahr10}
{Ormel} C.~W.,  {Klahr} H.~H.,  2010, \mn@doi [\aap]
  {10.1051/0004-6361/201014903}, \href
  {http://adsabs.harvard.edu/abs/2010A%26A...520A..43O} {520, A43}

\bibitem[\protect\citeauthoryear{{Ormel}, {Kuiper}  \& {Shi}}{{Ormel}
  et~al.}{2015a}]{ormel1}
{Ormel} C.~W.,  {Kuiper} R.,   {Shi} J.-M.,  2015a, \mn@doi [\mnras]
  {10.1093/mnras/stu2101}, \href
  {http://adsabs.harvard.edu/abs/2015MNRAS.446.1026O} {446, 1026}

\bibitem[\protect\citeauthoryear{{Ormel}, {Shi}  \& {Kuiper}}{{Ormel}
  et~al.}{2015b}]{ormel2}
{Ormel} C.~W.,  {Shi} J.-M.,   {Kuiper} R.,  2015b, \mn@doi [\mnras]
  {10.1093/mnras/stu2704}, \href
  {http://adsabs.harvard.edu/abs/2015MNRAS.447.3512O} {447, 3512}

\bibitem[\protect\citeauthoryear{{Owen} \& {Jackson}}{{Owen} \&
  {Jackson}}{2012}]{owenjackson12}
{Owen} J.~E.,  {Jackson} A.~P.,  2012, \mn@doi [\mnras]
  {10.1111/j.1365-2966.2012.21481.x}, \href
  {http://adsabs.harvard.edu/abs/2012MNRAS.425.2931O} {425, 2931}

\bibitem[\protect\citeauthoryear{{Owen} \& {Wu}}{{Owen} \&
  {Wu}}{2013}]{owenwu13}
{Owen} J.~E.,  {Wu} Y.,  2013, \mn@doi [\apj] {10.1088/0004-637X/775/2/105},
  \href {http://adsabs.harvard.edu/abs/2013ApJ...775..105O} {775, 105}

\bibitem[\protect\citeauthoryear{{Paardekooper} \& {Papaloizou}}{{Paardekooper}
  \& {Papaloizou}}{2009}]{PP09}
{Paardekooper} S.-J.,  {Papaloizou} J.~C.~B.,  2009, \mn@doi [\mnras]
  {10.1111/j.1365-2966.2009.14511.x}, \href
  {http://adsabs.harvard.edu/abs/2009MNRAS.394.2283P} {394, 2283}

\bibitem[\protect\citeauthoryear{{Perri} \& {Cameron}}{{Perri} \&
  {Cameron}}{1974}]{perricameron74}
{Perri} F.,  {Cameron} A.~G.~W.,  1974, \mn@doi [\icarus]
  {10.1016/0019-1035(74)90074-8}, \href
  {http://adsabs.harvard.edu/abs/1974Icar...22..416P} {22, 416}

\bibitem[\protect\citeauthoryear{{Plummer}}{{Plummer}}{1911}]{plummer11}
{Plummer} H.~C.,  1911, \mn@doi [\mnras] {10.1093/mnras/71.5.460}, \href
  {http://adsabs.harvard.edu/abs/1911MNRAS..71..460P} {71, 460}

\bibitem[\protect\citeauthoryear{{Pollack}, {Hubickyj}, {Bodenheimer},
  {Lissauer}, {Podolak}  \& {Greenzweig}}{{Pollack} et~al.}{1996}]{pollack96}
{Pollack} J.~B.,  {Hubickyj} O.,  {Bodenheimer} P.,  {Lissauer} J.~J.,
  {Podolak} M.,   {Greenzweig} Y.,  1996, \mn@doi [\icarus]
  {10.1006/icar.1996.0190}, \href
  {http://adsabs.harvard.edu/abs/1996Icar..124...62P} {124, 62}

\bibitem[\protect\citeauthoryear{{Rafikov}}{{Rafikov}}{2006}]{rafikov06}
{Rafikov} R.~R.,  2006, \mn@doi [\apj] {10.1086/505695}, \href
  {http://adsabs.harvard.edu/abs/2006ApJ...648..666R} {648, 666}

\bibitem[\protect\citeauthoryear{{Rafikov}}{{Rafikov}}{2011}]{rafikov11}
{Rafikov} R.~R.,  2011, \mn@doi [\apj] {10.1088/0004-637X/727/2/86}, \href
  {http://adsabs.harvard.edu/abs/2011ApJ...727...86R} {727, 86}

\bibitem[\protect\citeauthoryear{{Rafikov}}{{Rafikov}}{2016}]{rafikov16}
{Rafikov} R.~R.,  2016, \mn@doi [\apj] {10.3847/0004-637X/831/2/122}, \href
  {http://adsabs.harvard.edu/abs/2016ApJ...831..122R} {831, 122}

\bibitem[\protect\citeauthoryear{Rivier, Crida, Morbidelli  \& Brouet}{Rivier
  et~al.}{2012}]{rivier12}
Rivier G.,  Crida A.,  Morbidelli A.,   Brouet Y.,  2012, \mn@doi [Astronomy \&
  Astrophysics] {10.1051/0004-6361/201218879}, 548, A116

\bibitem[\protect\citeauthoryear{Roe}{Roe}{1981}]{roe81}
Roe P.,  1981, \mn@doi [Journal of Computational Physics]
  {https://doi.org/10.1016/0021-9991(81)90128-5}, 43, 357

\bibitem[\protect\citeauthoryear{{Rogers}}{{Rogers}}{2015}]{rogers15}
{Rogers} L.~A.,  2015, \mn@doi [\apj] {10.1088/0004-637X/801/1/41}, \href
  {http://adsabs.harvard.edu/abs/2015ApJ...801...41R} {801, 41}

\bibitem[\protect\citeauthoryear{Safronov}{Safronov}{1969}]{safronov69}
Safronov V.~S.,  1969, NASA Tech. Trans., pp F--677

\bibitem[\protect\citeauthoryear{Szul\'agyi, Masset, Lega, Crida, Morbidelli
  \& Guillot}{Szul\'agyi et~al.}{2016}]{szulagyi16}
Szul\'agyi J.,  Masset F.,  Lega E.,  Crida A.,  Morbidelli A.,   Guillot T.,
  2016, \mn@doi [Monthly Notices of the Royal Astronomical Society]
  {10.1093/mnras/stw1160}, 460, 2853

\bibitem[\protect\citeauthoryear{{Tanigawa}, {Ohtsuki}  \&
  {Machida}}{{Tanigawa} et~al.}{2012}]{tanigawa12}
{Tanigawa} T.,  {Ohtsuki} K.,   {Machida} M.~N.,  2012, \mn@doi [\apj]
  {10.1088/0004-637X/747/1/47}, \href
  {http://adsabs.harvard.edu/abs/2012ApJ...747...47T} {747, 47}

\bibitem[\protect\citeauthoryear{Van~Leer}{Van~Leer}{}]{van1997relation}
Van~Leer B., , in Hussaini M.~Y.,  Van~Leer B.,   Van~Rosendale J.,  eds, ,
  Upwind and high-Resolution schemes.
Springer Berlin Heidelberg, pp 33--52

\bibitem[\protect\citeauthoryear{Van~Leer}{Van~Leer}{1979}]{vanleer79}
Van~Leer B.,  1979, Journal of Computational Physics, 32, 101

\bibitem[\protect\citeauthoryear{{Ward}}{{Ward}}{1991}]{ward91}
{Ward} W.~R.,  1991, in Proceedings of Lunar and Planetary Science Conference.
  Lunar and Planetary institute, Houson, TX, p.~80

\bibitem[\protect\citeauthoryear{{Weidenschilling}}{{Weidenschilling}}{1977}]{weidenschilling77}
{Weidenschilling} S.~J.,  1977, \mn@doi [\mnras] {10.1093/mnras/180.1.57},
  \href {http://adsabs.harvard.edu/abs/1977MNRAS.180...57W} {180, 57}

\bibitem[\protect\citeauthoryear{{Weiss} \& {Marcy}}{{Weiss} \&
  {Marcy}}{2014}]{weissmarcy14}
{Weiss} L.~M.,  {Marcy} G.~W.,  2014, \mn@doi [\apjl]
  {10.1088/2041-8205/783/1/L6}, \href
  {http://adsabs.harvard.edu/abs/2014ApJ...783L...6W} {783, L6}

\bibitem[\protect\citeauthoryear{{Wetherill} \& {Stewart}}{{Wetherill} \&
  {Stewart}}{1989}]{wetherill89}
{Wetherill} G.~W.,  {Stewart} G.~R.,  1989, \mn@doi [\icarus]
  {10.1016/0019-1035(89)90093-6}, \href
  {http://adsabs.harvard.edu/abs/1989Icar...77..330W} {77, 330}

\bibitem[\protect\citeauthoryear{{Wolfgang} \& {Lopez}}{{Wolfgang} \&
  {Lopez}}{2015}]{wolfganglopez15}
{Wolfgang} A.,  {Lopez} E.,  2015, \mn@doi [\apj]
  {10.1088/0004-637X/806/2/183}, \href
  {http://adsabs.harvard.edu/abs/2015ApJ...806..183W} {806, 183}

\bibitem[\protect\citeauthoryear{{Yalinewich} \& {Schlichting}}{{Yalinewich} \&
  {Schlichting}}{2018}]{Yalin18}
{Yalinewich} A.,  {Schlichting} H.~E.,  2018, arXiv e-prints, \href
  {http://adsabs.harvard.edu/abs/2018arXiv181111778Y} {}

\bibitem[\protect\citeauthoryear{Yang \& Brent}{Yang \& Brent}{2002}]{KSPIBCGS}
Yang L.~T.,  Brent R.,  2002, in Proceedings of the Fifth International
  Conference on Algorithms and Architectures for Parallel Processing.

\makeatother
\end{thebibliography}

%%%%%%%%%%%%%%%%%%%%%%%%%%%%%%%%%%%%%%%%%%%%%%%%%%

%%%%%%%%%%%%%%%%% APPENDICES %%%%%%%%%%%%%%%%%%%%%

\appendix

\section{Convergence study} \label{app:convstudy}

We performed a series of isothermal 2D simulations with $H=16$ and $B=8$ to evaluate the spatial resolution required for convergence of most diagnostics. The radial domain $r/r_c \in \left[1,128\right]$ is meshed by either $64$, $128$, $256$ or $512$ logarithmically spaced cells.

\begin{figure}%[H]
  \begin{center}
    \includegraphics[width=\columnwidth]{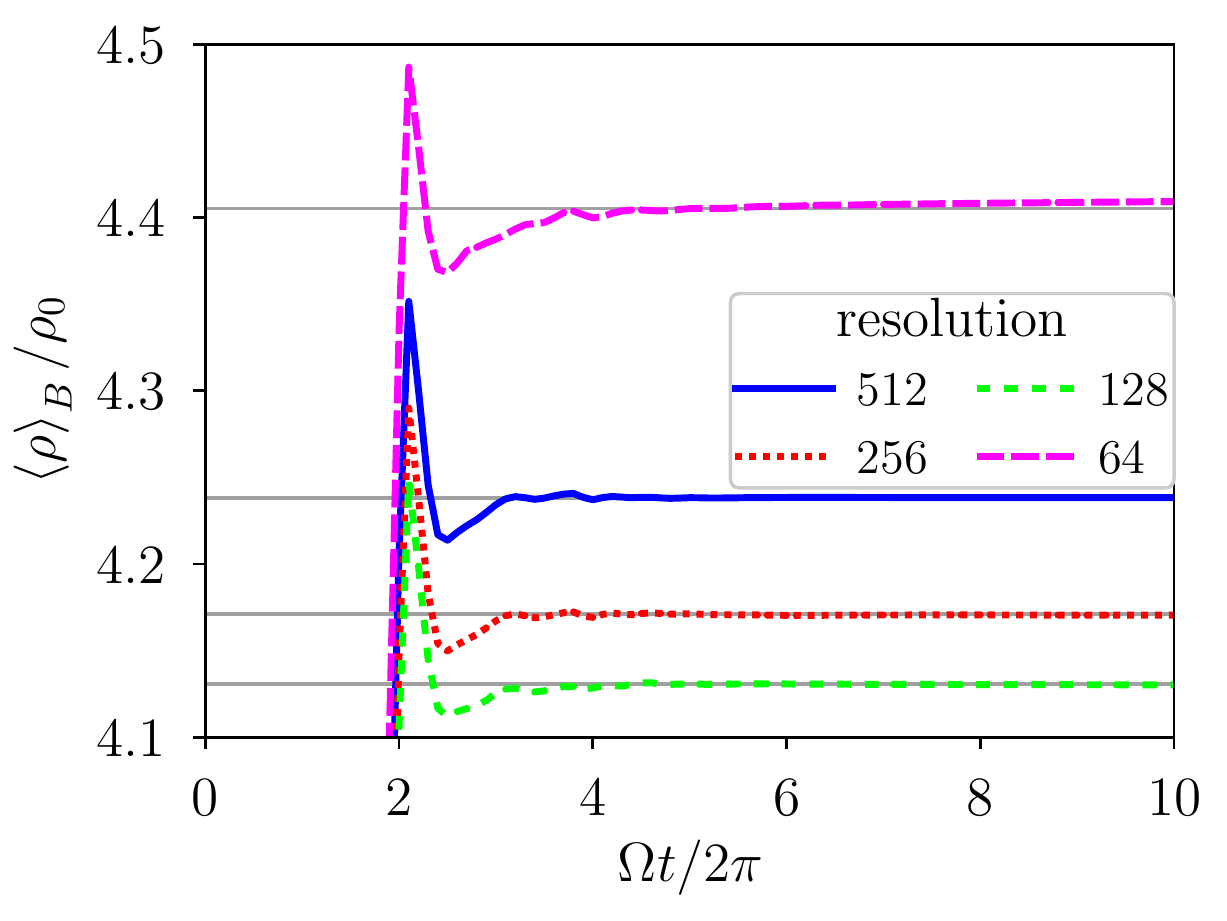}
    \caption{Surface-averaged density in the Bondi disk as a function of time for different resolutions; the first two orbits are used to progressively introduce the gravitational potential of the core. \label{fig:res_bmass}}
  \end{center}
\end{figure}

\autoref{fig:res_bmass} shows the temporal evolution of the surface-averaged density inside the Bondi disk. The average density increases during the first two orbits as the potential of the core is progressively introduced. It takes two more orbits for the envelope to adjust to the final potential and reach a quasi-steady state. The envelope mass varies by less than $10\%$ depending on the spatial resolution. After that, the envelope mass evolves on much longer timescales: the doubling time ranges from $10^4$ to over $10^5$ orbits from the lowest to the highest resolution. It corresponds to the timescale of kinetic energy dissipation by numerical viscosity in the envelope. 

\begin{figure}%[H]
  \begin{center}
    \includegraphics[width=\columnwidth]{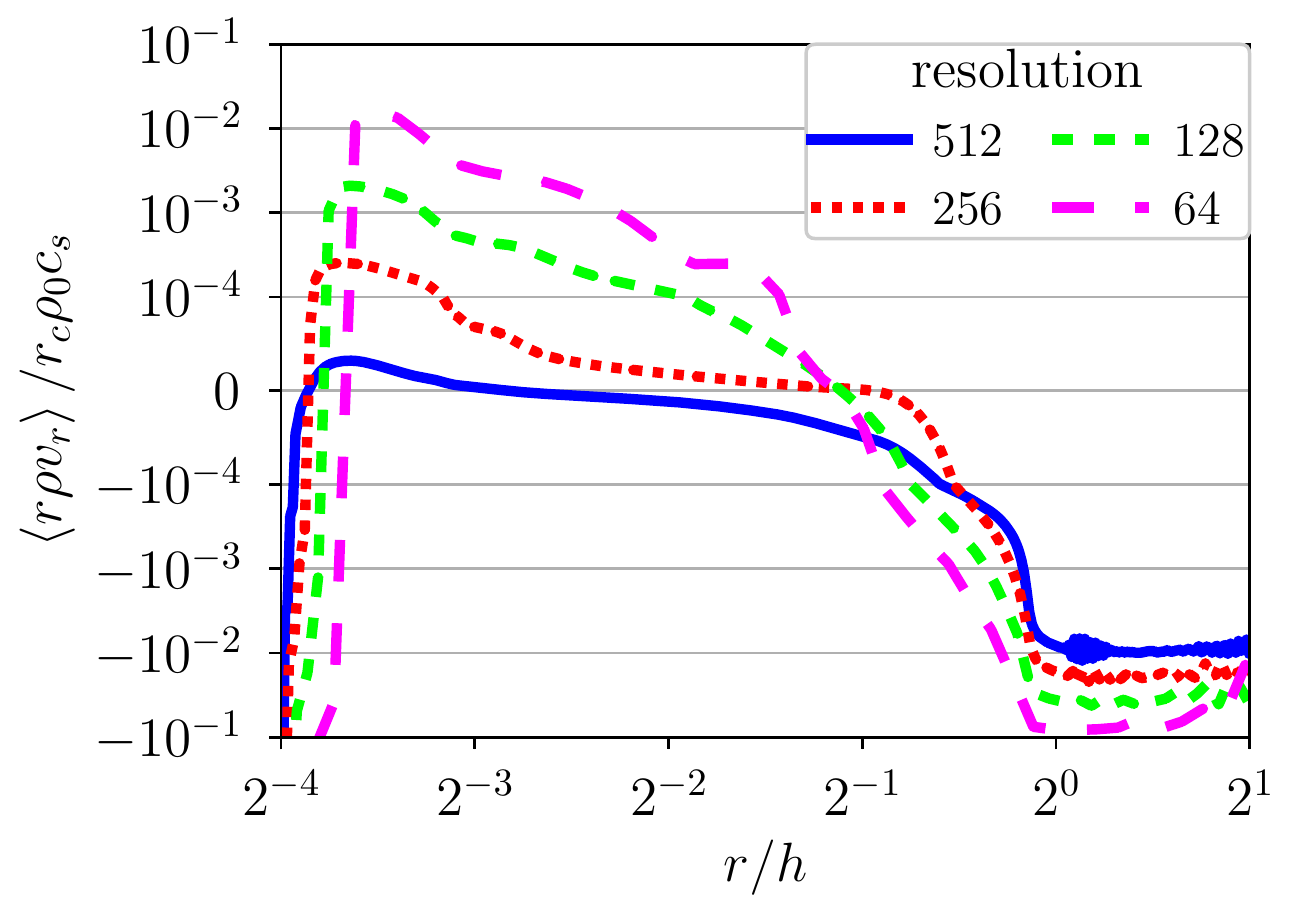}
    \caption{Time and azimuthally-averaged radial mass flux in the same series of resolution tests as on \autoref{fig:res_bmass}. \label{fig:res_mFr}}
  \end{center}
\end{figure}

On \autoref{fig:res_mFr}, we examine how the mass accretion flux depends on the grid resolution. This flux is estimated from the time and cell-averaged primitive variables $(\rho,v_r)$. There is a clear convergence with resolution toward a flat profile with zero mass flux in the envelope: the mass flux in the envelope goes down by almost a factor ten every time the resolution is doubled. We choose to use $512$ radial cells to make such mass fluxes insignificant. The measured mass flux is oriented inward at the core surface, outward in the envelope and inward out of the envelope. This suggests mass depletion near the core surface and accumulation at the envelope boundary. In reality, the inner radial boundary condition cancels the mass flux through the core surface. 

We emphasize that the \textsc{pluto} code uses a reconstruction scheme to estimate the primitive variables at the cell interfaces before converting to conservative variables such as the mass flux. Unfortunately, the mass flux actually used by the code is not easily accessible during computations. To confirm the absence of mass losses through the inner radial boundary, we measured the increase rate of the mass contained inside $r<h$. Independently, we averaged the integrated mass flux $2\pi r \rho v_r$ inside $r/h\in[1/4,1]$. These two diagnostics match to $10^{-2}$ accuracy when there is a net mass flux through the envelope as in \autoref{fig:i2dp16b16_SH_masstorques} or \autoref{fig:i2dp16b32_SH_massflux}. 

\section{Self-gravity in \textsc{pluto}} \label{app:selfgravity}

\subsection{Implementation}

The gravitational potential $\Phi_g$ satisfies Poisson's equation
\begin{equation} \label{eqn:poisson}
  \Delta \Phi_g = 4\pi \rho
\end{equation}
in the entire computational domain with appropriate boundary conditions. The structures and routines used for the representation and resolution of the parallel problem of solving this equation come from the PETSc library \citep{petsc-efficient,petsc-user-ref}. The Laplacian operator is discretized via second order finite difference. Boundary conditions are applied in the appropriate rows of the operator matrix. Given a density distribution, the linear problem for $\Phi_c$ is solved iteratively via a biconjugate gradient method \citep{KSPIBCGS}. By construction, the potential $\Phi_g$ is defined at cell centers; the interface values used by the hydrodynamic solver are computed by linear interpolation. Note that the boundary conditions are applied in the first active cell of the domain, and not at the boundary nor in the ghost zones. 

We implemented and tested this Poisson solver in cartesian, cylindrical and spherical geometries as described in the following sections. In the case of two-dimensional cylindrical geometry, the Green's function of the Laplacian operator is the potential of an infinite line $\Phi_g(r) \sim \log(r)$. Instead, we use the spherical representation of the Laplacian operator over the cylindrical coordinates $(r,\varphi)$, so that the potential of a point mass varies as the three-dimensional Newtonian potential $\sim 1/r$. 

The boundary conditions in the angular directions respect the topology of the domain (periodic in the azimuthal angle $\varphi$). The potential is set to $\Phi_g=0$ at the outer radial boundary, so as to fix a reference value. By doing so, we enforce an axial symmetry of the gas potential at large distances from the core. We keep this choice for simplicity, since we cannot know a priori the boundary values $\Phi_g\left(r_{\mathrm{out}},\varphi\right)$ for non-axisymmetric density distributions. The condition $\partial_r \Phi_g=0$ is imposed at the inner radial boundary, consistent with the absence of gas below the core radius $r_c$. Any other choice would also favor an unwanted mass flux through the inner boundary.

The parallel Poisson solver can significantly affect the performances of the code depending on the spatial resolution and dimensionality of the problem. In 3D spherical geometry, with a resolution of $[128\times 80 \times 160]$ on $(r,\theta,\varphi) \in \left[1,128\right]\times\left[0,\pi\right]\times\left[0,2\pi\right]$, the Poisson solver takes the equivalent of $2.3$ hydrodynamic timesteps to determine $\Phi_g$ for a constant density distribution and Dirichlet radial boundary conditions. This computational overhead can reach a factor $25$ in high-resolution cylindrical (2D) simulations. Since the large-scale structures of the density distribution do not evolve on short timescales near the core, it is reasonable to solve \eqref{eqn:poisson} for $\Phi_g$ every $n=4$ hydrodynamic timesteps, which we opted to do in our simulations (see \autoref{sec:fisherjeans}).

\subsection{Static tests}

The Poisson solver was tested in static configurations for cartesian, cylindrical and spherical geometries, in two and three dimensions. In these static tests, the mass distribution $\rho$ is not allowed to evolve in time. Simple prescriptions for the source term $\rho$ allow comparing $\Phi_g$ with analytical solutions of Poisson's equation. 

\begin{figure}%[H]
\begin{center}
\includegraphics[width=\columnwidth]{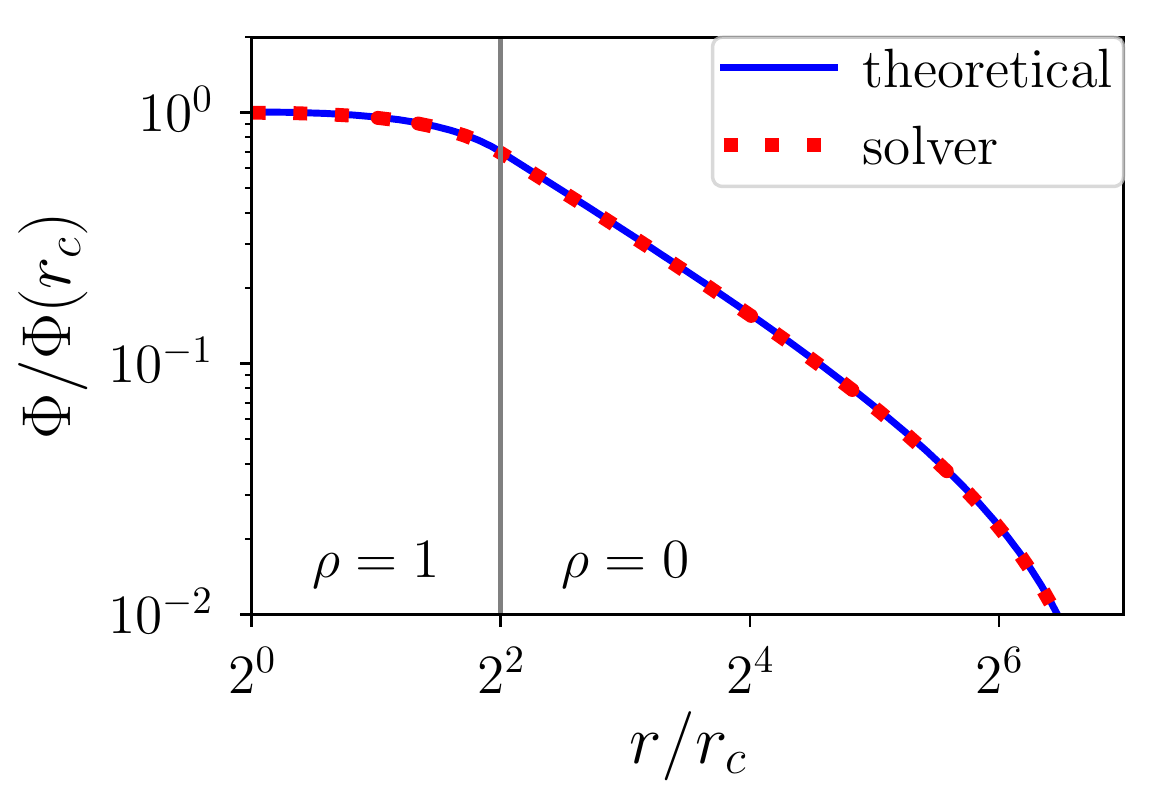}
\caption{Absolute value of the gravitational potential obtained from the Poisson solver (solid blue) and theoretical solution (dashed green) in our standard cylindrical setup containing a cylinder $\rho(r)=1$ inside $r/r_c<4$.
\label{fig:fisherstatcyl}}
\end{center}
\end{figure}

In 1D, imposing a constant density $\rho=1$ on the interval $x\in \left[0,1\right]$ with homogeneous Dirichlet conditions, the exact solution is $\Phi(x)=2\pi x(x-1)$. Because the boundary conditions are imposed at the center of the first active cells and not at their boundaries, the approximate $\Phi_g$ is offset with respect to the exact one by a constant. After subtracting this constant, the residual error oscillates between $\pm 10^{-7}$, so the first and second derivatives of $\Phi_g$ are accurately captured. 

We perform a similar test within our standard cylindrical setup by prescribing $\rho(r)=1$ for $r<4 r_c$ and $\rho(r)=0$ beyond. The numerical and theoretical profiles of $\vert\Phi_g(r)\vert$ are in excellent agreement as shown on \autoref{fig:fisherstatcyl}. The relative error on $\Phi_g$ is less than $1\%$ inside $r\leq 32$ and reaches $2\%$ at $r=92r_c$. 

To test the proper implementation of the boundary conditions, additional static tests were performed with an asymmetric source term. For fully periodic domains, the compatibility condition\footnote{In Fourier space, the zeroth-order (constant) component of \eqref{eqn:poisson} must reduce to zero.} $\int \rho = 0$ over the whole domain is enforced by subtracting the volume-averaged density in \eqref{eqn:poisson}. 

\subsection{Dynamic test: Jeans instability} \label{sec:fisherjeans}

We verified that our implementation performs well in dynamic situations by reproducing Jeans' instability in Cartesian geometry with periodic boundary conditions. The domain $\left(x,y\right) \in \left[0,1\right]\times\left[0,2\right]$ is meshed with $64\times 128$ cells, the fluid is initialized with zero velocity, and the density $\rho = 1+\epsilon$ is flat with a white noise of amplitude $\epsilon=10^{-6}$. The Jeans length is set to $\lJ=3/2$, so harmonic perturbations are unstable only in the $y$ direction. 

\begin{figure}%[H]
\begin{center}
\includegraphics[width=\columnwidth]{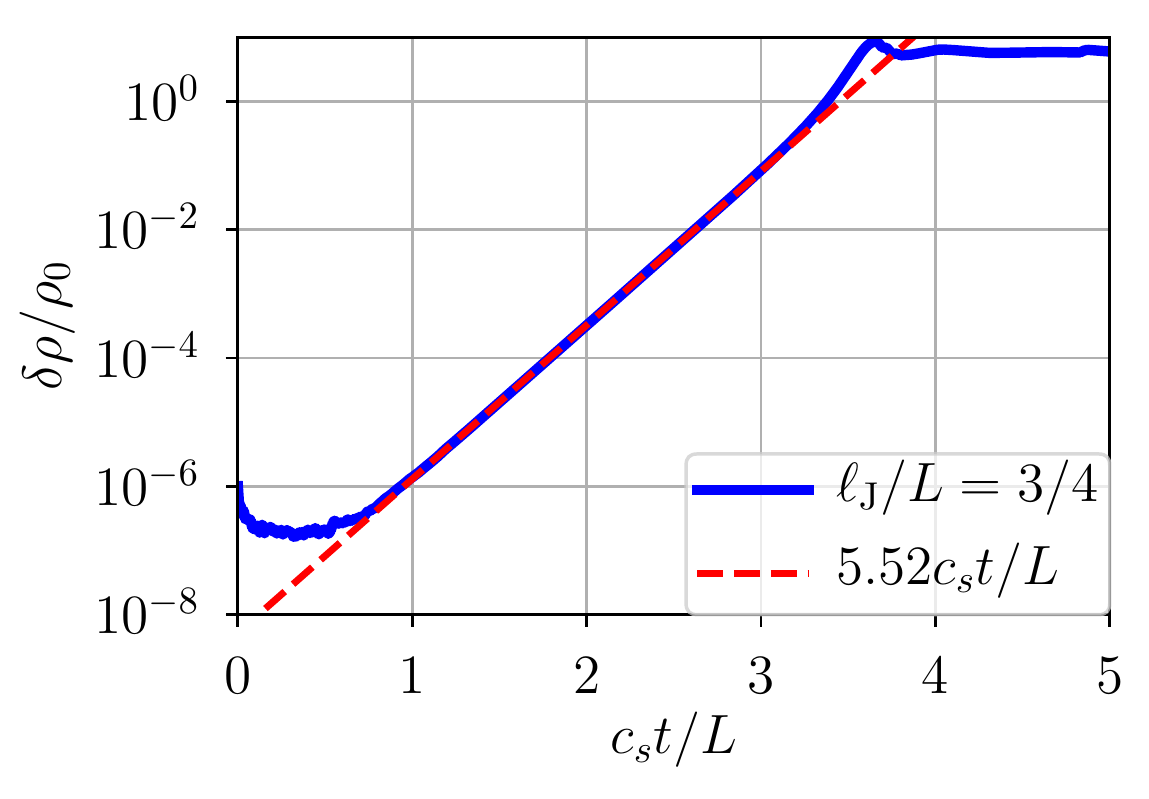}
\caption{Amplitude of density fluctuations following Jeans' instability for $\lJ/L=3/4$ when solving Poisson's equation \eqref{eqn:poisson} at every hydrodynamic timestep; the analytical prediction for the linear phase (dashed red) matches the simulation data (solid blue) to $4\times 10^{-3}$ accuracy. \label{fig:fisherjeans}}
\end{center}
\end{figure}

We start by computing $\Phi_g$ at every timestep of the Runge-Kutta 2 integrator (every two substeps). The linear mode with length $L=2$ has an expected growth rate $s = 2\pi  \left(c_s/L\right) \sqrt{\vert1-\left(L/\lJ\right)^2\vert} \approx 5.54 \left(c_s/L\right)$. As demonstrated on \autoref{fig:fisherjeans}, the measured growth rate $5.520\pm 0.001$ matches the theoretical one to $4\times 10^{-3}$ accuracy in this configuration. We reproduced Jeans' instability by solving \eqref{eqn:poisson} every $n=4$ and $n=10$ timesteps, and measured relative errors of $2\%$ and $5\%$ on the growth rates respectively.

%%%%%%%%%%%%%%%%%%%%%%%%%%%%%%%%%%%%%%%%%%%%%%%%%%

% Don't change these lines
\bsp	% typesetting comment
\label{lastpage}
\end{document}